\documentclass[10pt,journal,compsoc]{IEEEtran}

\usepackage{url}
\usepackage{float}
\usepackage{csquotes}
\usepackage{xcolor}
\usepackage{xcolor}
\usepackage{graphicx}
\usepackage{url}
\usepackage{subcaption}

\ifCLASSOPTIONcompsoc
  \usepackage[nocompress]{cite}
\else
  \usepackage{cite}
\fi
\ifCLASSINFOpdf
\else
\fi
\begin{document}
%
\title{\huge AI-native Interconnect Framework for Integration of Large Language Model Technologies in 6G Systems}
%
%
%
%

\author{Sasu~Tarkoma$^*$,
        Roberto~Morabito$^*$,
        Jaakko~Sauvola
        \\[12pt]
        
\IEEEcompsocitemizethanks{\IEEEcompsocthanksitem Sasu Tarkoma (e-mail: sasu.tarkoma@helsinki.fi) and Roberto Morabito (e-mail: roberto.morabito@helsinki.fi) are with the Department of Computer Science, University of Helsinki (Finland). Jaakko Sauvola (e-mail: jaakko.sauvola@oulu.fi) is with the Department of Electrical and Information Engineering, University of Oulu (Finland). 
\newline
* These authors contributed equally to this work.\protect\\
}
\thanks{This paper is currently under review for potential inclusion in the series of White Papers of the 6G Flagship research program (https://www.6gflagship.com/white-papers/). Please be aware that the content may be subject to further revisions or edits before its final publication there. Updated versions of this document may be released in the future.
\newline
\newline
We extend our sincere gratitude to all other contributors for their valuable comments and insights, which have significantly enhanced the quality of this paper. Special thanks to Mehdi Bennis, Abhishek Kumar, Lauri Loven, Jukka Riekki, Sumudu Samarakoon (University of Oulu), Naser Hossein Motlagh (University of Helsinki), Leandros Tassiulas (Yale University), Ian Akyildiz (Truva Inc.), Aaron Ding (TU Delft), for their dedicated input and feedback.
}
}

\IEEEtitleabstractindextext{%
\begin{abstract}
The evolution towards 6G architecture promises a transformative shift in communication networks, with artificial intelligence (AI) playing a pivotal role. This paper delves deep into the seamless integration of Large Language Models (LLMs) and Generalized Pretrained Transformers (GPT) within 6G systems. Their ability to grasp intent, strategize, and execute intricate commands will be pivotal in redefining network functionalities and interactions. Central to this is the AI Interconnect framework, intricately woven to facilitate AI-centric operations within the network.
Building on the continuously evolving current state-of-the-art, we present a new architectural perspective for the upcoming generation of mobile networks. Here, LLMs and GPTs will collaboratively take center stage alongside traditional pre-generative AI and machine learning (ML) algorithms. This union promises a novel confluence of the old and new, melding tried-and-tested methods with transformative AI technologies. Along with providing a conceptual overview of this evolution, we delve into the nuances of practical applications arising from such an integration. Through this paper, we envisage a symbiotic integration where AI becomes the cornerstone of the next-generation communication paradigm, offering insights into the structural and functional facets of an AI-native 6G network.
\end{abstract}

\begin{IEEEkeywords}
6G, Generative AI, Large Language Model (LLM), Generative Pre-trained Transformers (GPT), AI Interconnect, Edge Intelligence, Open RAN (O-RAN).
\end{IEEEkeywords}}

\maketitle

\IEEEdisplaynontitleabstractindextext

%
\IEEEpeerreviewmaketitle

\ifCLASSOPTIONcompsoc
\IEEEraisesectionheading{\section{Introduction}\label{sec:introduction}}
\else
\section{Introduction}
\label{sec:introduction}
\fi
The transformational potential of AI is undeniably evident across various sectors, particularly within the telecommunications industry. As AI's role in telecom use cases has grown in recent years, the phrase \enquote{AI-native telco} has gained traction. Numerous recent contributions from both industry and academia, which delineate the path for the next generation of mobile networks and define their requirements, roadmaps, and visions for the 6G architecture \cite{britto2023telecom, ainative_nokia, flagshipwp}, describe an AI-native system as one inherently endowed with trustworthy AI capabilities. Such systems seamlessly integrate AI into their design, deployment, operation, and maintenance. A hallmark of these AI-native systems is their data-centric, knowledge-driven environment. In this setting, data is continuously produced and utilized to unveil novel AI-driven functionalities, enabling the support of intelligent applications across a multitude of heterogeneous use cases and domains \cite{latva2019key, uusitalo20216g}. This represents a significant departure from conventional static, rule-based systems, transitioning towards more adaptive, learning-oriented AI models as needed.
Large Language Models (LLMs), and more specifically Generalized Pretrained Transformers (GPT) \cite{bommasani2021opportunities, zhao2023survey, brown2020language}, have become very powerful tools for understanding intent, creating plans and tools, and formulating and executing complex instructions \cite{zhao2023survey}. This capability is significantly enhanced by prompt engineering, the means by which LLMs are programmed via prompts \cite{white2023prompt}. 
As such, they are essential building blocks of future intelligent networks and applications. 

In this work, we advocate for an AI-native 6G network that seamlessly integrates a diverse range of LLMs, allowing for their dynamic selection, provisioning, updating, and creation. Central to this vision is the \textbf{AI Interconnect}, which streamlines AI-centric operations within the network. 
By incorporating AI directly into the network, we anticipate marked improvements in areas such as radio and network optimization, privacy and security via tailored AI and resource selection, enhanced accountability through meticulous AI operation monitoring and audit trails, and meeting stringent latency and other network-specific criteria. Such advancements are poised to be invaluable for domains like the Internet of Things (IoT), robotics, smart cities, and autonomous systems, to name a few.

\subsection*{Motivation}
In June 2023, the International Telecommunication Union's Radio Communication Sector Working Party (ITU-R WP 5D) marked a significant milestone towards the advent of 6G by approving the Framework Recommendation for IMT-2030 \cite{ITU-R-IMT-2030}. This pivotal framework, represented through a wheel-shaped diagram, delineates six crucial usage scenarios, underpinned by four overarching design principles. These principles comprise \emph{(i)} \textit{sustainability}, \emph{(ii)} \textit{security, privacy, and resilience}, \emph{(iii)} \textit{connecting the unconnected}, and \emph{(iv)} \textit{ubiquitous intelligence}.

The inclusion of \textit{AI and Communication} among the six key usage scenarios, combined with the emphasis on \textit{ubiquitous intelligence}, underscores that 6G's promise extends beyond mere improvements in speed and reliability. It predicts a paradigm wherein AI-driven autonomous systems, dynamic network configurations, and edge intelligence become the standard. This  trajectory is exemplified by the concept of \enquote{AI-native Telecom}, where AI transitions from being an adjunct feature to the core of both hardware and software network components \cite{ainative}. Such synergy is anticipated to foster unprecedented enhancements in network efficiency and adaptability. Moreover, the consolidation of AI with 6G is ready to enable a new set of applications and services — from ultra-reliable low-latency communication to immersive augmented realities — reaffirming AI's essential role in the forthcoming communication era \cite{bommasani2021opportunities, zhao2023survey, brown2020language, xu2021edge}.

However, charting the course to this AI-enhanced networked future is not without compelling challenges. The intricate nature of telecom and network ecosystems introduces considerable complexity, especially concerning cross-layer and seamless interoperability amongst various AI-enabled components. Despite these hurdles, the pursuit of innovative AI solutions, including foundational models like GPTs \cite{bommasani2021opportunities}, continues relentlessly. Such a transformation mandates a rethinking of our traditional engineering approaches, propelling us into a new era characterized by data-centric, knowledge-driven ecosystems \cite{tataria20216g}. In this evolving landscape, LLMs and GPTs can make a significant impact, especially as we navigate the intricacies of 6G -- a domain where the stringent demands of connectivity, capacity, latency, mobility, and reliability become increasingly arduous to satisfy without the prowess of AI \cite{tataria20216g}.

\subsection*{AI Interconnect: A Glimpse}
Building on this AI-centric paradigm shift in telecommunications, our proposed AI Interconnect framework aligns seamlessly with the vision of AI-native networks. It anticipates and addresses the requirements pertinent to distributed AI operations, such as prompt processing and the effective selection and execution of LLMs. This aligns with the broader objective of fostering trust in the operation of AI frameworks, particularly in the context of the 6G architecture. Moreover, as we move forward in this direction, it is essential to holistically consider the myriad engineering implications introduced by this evolution, ensuring that we are not just technologically equipped but also strategically positioned to leverage the transformative power of AI in telecommunications.

Context, elements, state machines, and feedback loops are indispensable in ensuring optimal network performance, resilience, and adaptability. Feedback loops, such as the MAPE-K (Monitor-Analyze-Plan-Execute over a shared Knowledge) model \cite{gheibi2021applying}, allow these network elements and components to continuously monitor their performance, analyze the collected data, plan for enhancements, and execute these plans, thereby facilitating self-regulation and adaptation. Specifically, the MAPE-K model serves as the backbone of the AI Interconnect managing system, guiding it through adaptive cycles responsive to the 6G network’s dynamic environment. This seamless integration fosters a symbiotic relationship where the MAPE-K framework supports and is enhanced by LLM and GPT technologies. The MAPE-K and similar models are envisioned to leverage LLMs for advanced capabilities. The ReAct (\enquote{Reason+Act}) model represents a transformative approach in leveraging LLMs for agent-based actions within a specified environment \cite{yao2022react}. At its essence, ReAct utilizes the LLM as a dynamic planner by prompting it to \enquote{think out loud}. This is achieved by presenting the LLM with a comprehensive textual overview encompassing the current environment, a defined objective, an array of potential actions, and a chronological record of preceding actions and observations. Upon processing this data, the LLM generates a sequence of contemplative thoughts before arriving at a specific action. Once determined, this action is seamlessly executed within the stipulated environment, thereby establishing the LLM as a passive analytical tool and an active participant capable of decision-making and subsequent action.

\begin{figure}[ht!]
    \centering
      \includegraphics[width=\columnwidth]{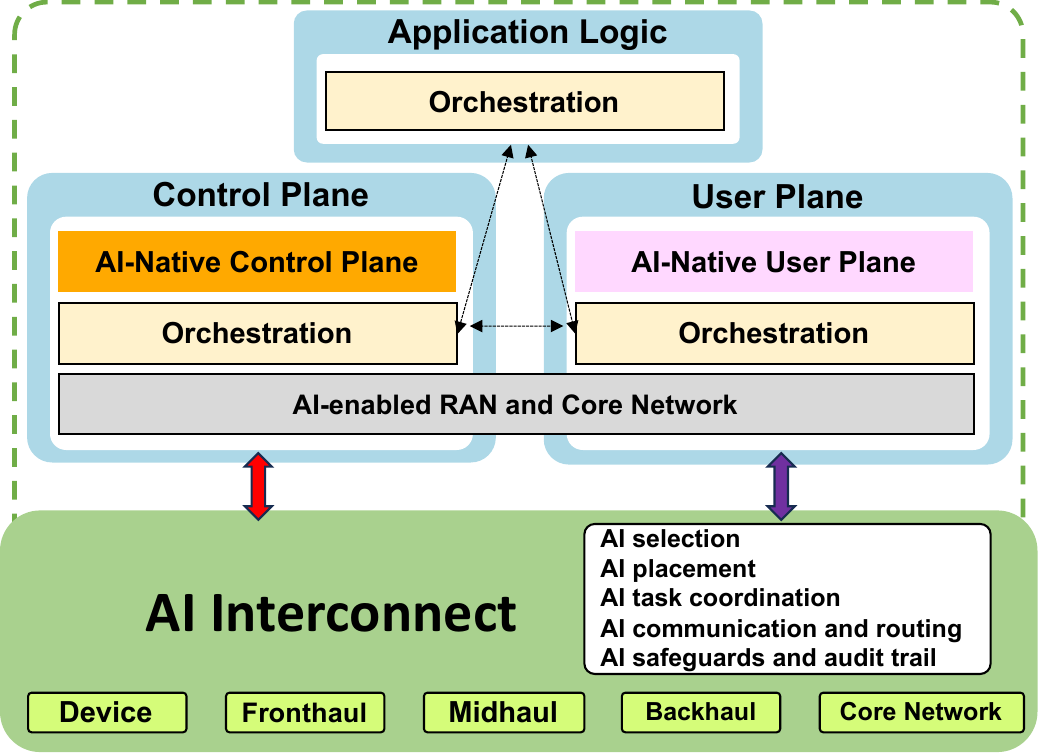}
        \centering
      \caption{AI Interconnect's cross-layer design across control, user, and application planes.}\label{fig:1}
\end{figure}

Figure \ref{fig:1} presents an overview of the AI Interconnect's cross-layer design, spanning the cloud-to-edge continuum and beyond. This design extends from devices to the core network, traversing fronthaul, midhaul, and backhaul which are collectively referred to as \textit{x-haul} \cite{townend2023challenges}. Fronthaul refers to the network connecting remote radio heads to BaseBand Units (BBUs) several kilometers away. Midhaul denotes the link between the Digital Unit (DU) and Centralized Unit (CU), while Backhaul represents the link connecting the CU to the core network \cite{isg20185g}.
The AI Interconnect is expected to accommodate learning and inference capabilities; however, in this paper, we focus on using high-level management and orchestrating LLM-style models and do not elaborate on specific distributed learning and inference solutions.
LLMs can help understand, manage, and coordinate network and application behaviours based on raw data, for example, KPIs and radio parameters, prior understanding of the network and the domains (expert models), and subscribing and publishing inference results through an interconnect. A set of expert models can support the overall orchestration and coordination of how the information and computing are realised in the edge-cloud continuum taking the communication, energy, privacy, security, sustainability, and computing constraints into account. Resource efficiency is an important design consideration for the AI Interconnect.

Bridging these high-level capabilities with their practical implementation, the AI Interconnect provides LLM messaging and brokering capabilities for the \textit{control plane}, \textit{user plane}, and higher-level application elements (i.e., \textit{Application Logic}). We envision the AI Interconnect to be \enquote{message-oriented}, featuring both \textit{request/reply} and \textit{publish/subscribe (pub/sub)} APIs for requesting AI inference and other learning capabilities supported by the network. The asynchronous message-based nature enables reactive tasks, and its message-oriented operation offers an auditing capability for LLM usage. Among the diverse functionalities of the AI Interconnect are AI selection, AI placement, AI task coordination, AI communication and routing, and AI safeguards with an audit trail. Given a task, the AI Interconnect assumes the responsibility of choosing appropriate LLMs to address the task, crafting a strategy to attain the task objective, and ensuring smooth communication between the LLMs and the network components. The message-centric nature of the Interconnect facilitates meticulous monitoring of AI pipelines' operations, a pivotal aspect for elucidating system behaviors and outcomes, and bolstering auditing processes.

Regarding support for 6G applications, the AI Interconnect aims to facilitate the use of LLMs for a wide range of application behaviors, including real-time content generation. For instance, edge servers can host both AI control logic, such as prompt engineering modules, and LLMs of different complexities. For applications, the AI Interconnect addresses the need for local and private placement of AI components, generating reports of AI operations for auditing, and is envisioned to accommodate a wide range of use cases. This includes general-purpose applications, network deployment and management, service-level agreements (SLAs) management, as well as evolved generative AI (GenAI) applications and application behaviours driven by user and application intent.

However, while the promise of LLMs for mobile and 6G networks is undeniably significant, it is important to note that they will function in tandem with traditional ML/AI models. Certain tasks within the telecom sphere may demand the granularity, transparency, or specificity that conventional ML/AI algorithms provide. LLMs may not be suitable for all tasks, especially those requiring in-depth analytical insights or real-time responsiveness. The AI Interconnect, with its robust architecture and cross-layer design, is particularly adept at accommodating this hybrid setup, seamlessly facilitating communication and cooperation between LLMs and traditional ML/AI models. Hence, a hybrid approach that leverages both LLMs and traditional ML/AI models within the AI Interconnect framework will likely be the most pragmatic solution.

\subsection*{Paper Structure}

Having highlighted the key premise of integrating LLMs and GPT technologies with 6G systems via a native AI Interconnect framework, we now proceed to introduce the structure of the remainder of this paper.

     \begin{figure}[htbp]
    \centering
      \includegraphics[width=1\columnwidth]{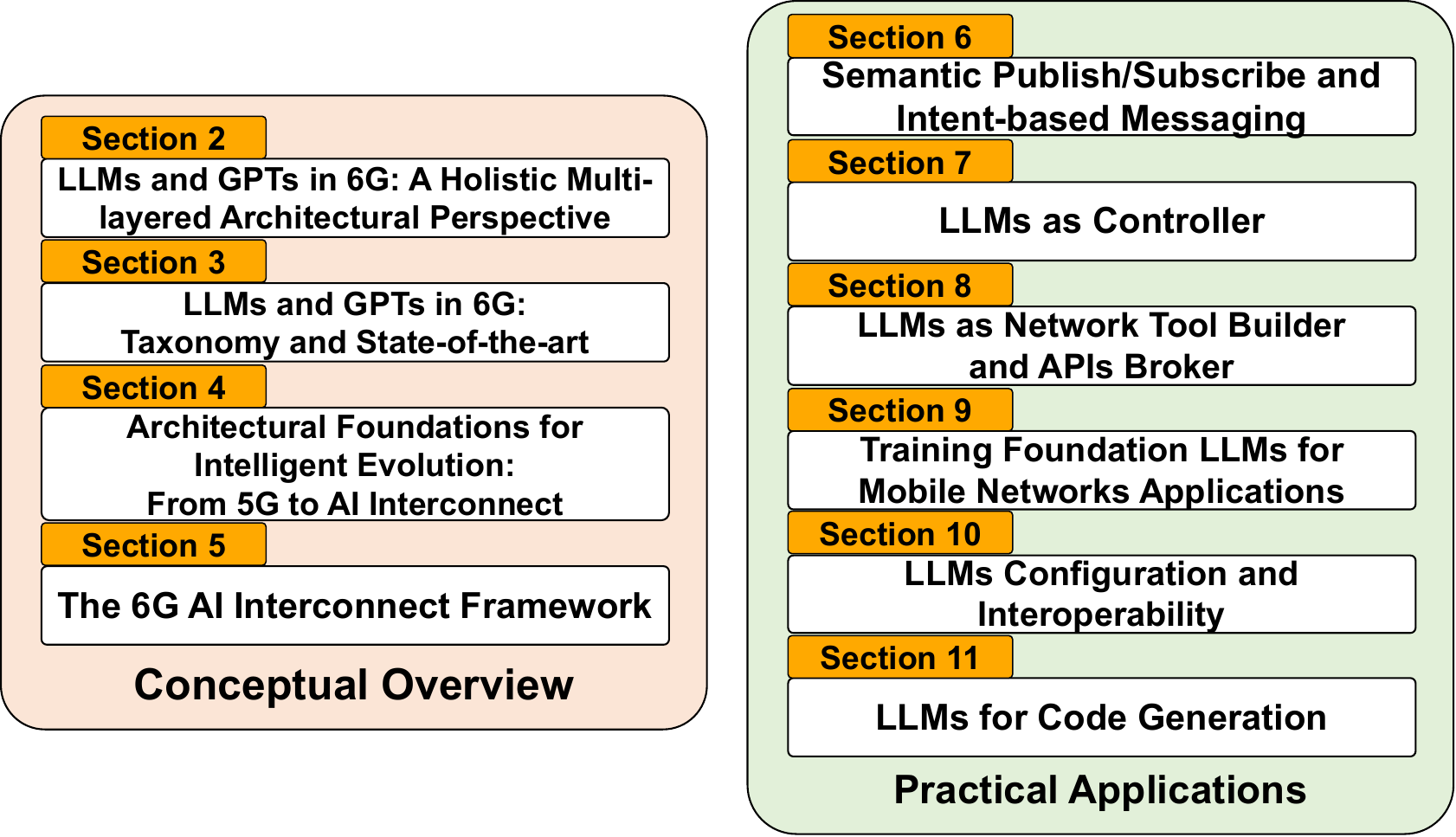}
       \centering
      \caption{Graphical representation of the paper's organization, highlighting the distinction between the 'Conceptual Overview' sections (2-5) and the 'Practical Applications' sections (6-11).}\label{paperstruct}
    \end{figure}

Initially, we delve into the \textbf{Conceptual Overview}, starting with Section \ref{sec:2} that provides a strategic architectural view of 6G integration with LLMs and GPT technologies. This leads into Section \ref{sec:3}, which offers a detailed taxonomy and state-of-the-art overview of LLMs and GPTs for 6G. Section \ref{sec:4} lays the groundwork for understanding the architectural foundations and the progressive integration of intelligence in 5G and beyond, setting the stage for the introduction of the 6G AI Interconnect framework discussed in Section \ref{sec:5}. Transitioning into the \textbf{Practical Applications}, Section \ref{sec:6} delves into the details of messaging methods used in the AI framework. In Section \ref{sec:8}, we explore the role of LLMs as controllers, and Section \ref{sec:9} elaborates on their functionality as dynamic tool builders and API brokers. Section \ref{sec:7} is dedicated to the specifics of training and fine-tuning LLMs for telecommunication applications. Section \ref{sec:10} discusses challenges such as GPT-ossification and provides practical insights into 6G-LLMs configuration and interoperability. Section \ref{sec:11} then discusses the role of LLMs in code generation, focusing particularly on dynamic processes.

     \begin{figure*}[ht!]
    \centering
      \includegraphics[width=0.8\textwidth]{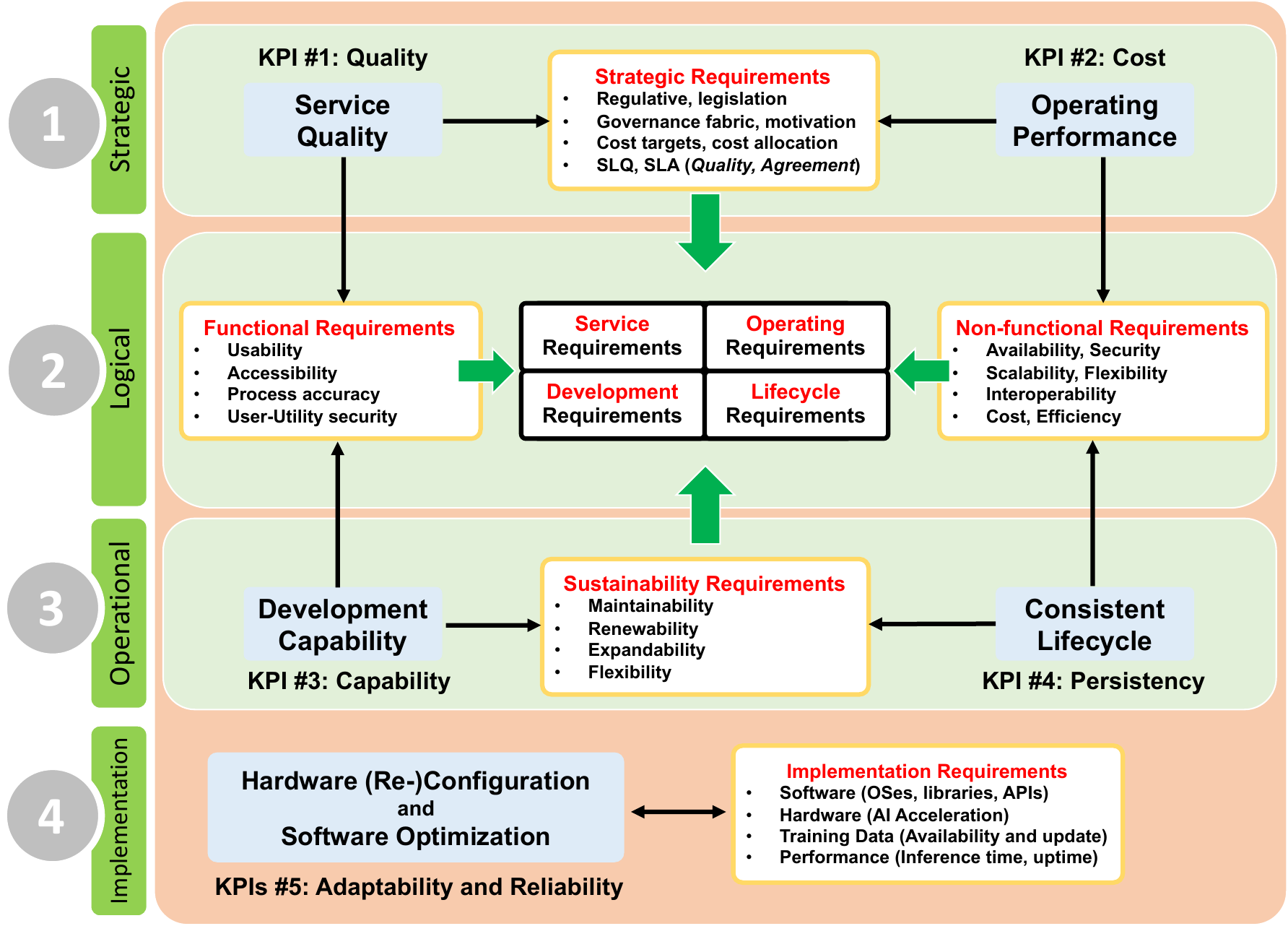}
       \centering
      {\textbf{\caption{A holistic representation of the 6G architectural framework, illustrating the layered approach from high-level strategic objectives to practical implementation considerations. This framework integrates LLM and GPT technologies across strategic, logical, operational, and implementation layers, with associated key performance indicators (KPIs) and requirements.}\label{fig:2}}}
    \end{figure*}
    
For a visual representation of how this paper is structured, readers can refer to Figure \ref{paperstruct} which depicts the delineation between the conceptual overview and practical applications within the paper.

\section{LLMs and GPTs in 6G: A Holistic Multi-layered Architectural Perspective}\label{sec:2}

The emergence of 6G with the integration of GPTs and LLMs has the potential to establish new standards for efficiency, flexibility, sustainability, and adaptability, even creating new service categories for different stakeholders in the entire 6G framework. According to our holistic view, the 6G architecture can be sectioned into four layers: \emph{(1)} \textit{strategic}, \emph{(2)} \textit{logical}, \emph{(3)} \textit{operational}, and \emph{(4)} \textit{implementation}. Within this framework, we need to deploy proper requirements for each layer, with attributes dealing with service quality, operating performance, development capability, consistent lifecycle management, and hardware configuration \& software optimization, each with their contributions to requirements of implementing the different variants of LLM and GPT technology models into 6G architecture, as depicted in Figure \ref{fig:2}. Such requirements can be clustered via the level of abstraction, purpose, and stakeholder role (e.g., management, architects, operators, developers, and users). 

In this framework, each layer serves specific functions in the 6G ecosystem, culminating in the implementation layer, which stands as the execution platform. In fact, the implementation layer stands out due to its tangible, practical nature and requirements, whereas the others provide a more high-level, theoretical framework. Being the most practical of the lot, this layer serves as the execution platform for the strategies outlined in the preceding layers.  In the upper layers, to understand the potential of adopting and adapting suitable LLM and GPT technology variants in 6G, we propose three key requirement clusters for each purpose: A) \textit{Strategic requirements} that serve the highest-level guidance and regulation to the entire 6G architecture, mainly discussing service quality and operating performance issues; B) \textit{Functional and Non-Functional requirements} that provide directions for the operational behaviour of LLM/GPT. Functional requirements dictate the \enquote{perceived experience of the users of utilities}, while Non-functional requirements encompass over 20 attributes related to \enquote{non-visible experience, maintenance, and quality guarantees}. These requirements support over-the-life-time end-to-end operations, configuration, and maintenance. The third cluster offers C) \textit{Sustainability requirements} addressing the strict stipulations 6G implementation needs to obey, ensuring a consistent life cycle and continuous development capability for the entire 6G fabric.

Highlighting their interplay, all the requirement sections contribute to the integration of LLM/GPT into 6G. Each offers unique roles and implementation levels with a functional purpose that adds significant value to operational performance targets in 6G. This supports the evolution of 6G architecture to be adaptive to situational contexts, proximity requirements, and end-user needs.

To harness the potential of LLM and GPT technology role, implementation, value, and suitability across the four distinct layers of the 6G architecture, it is imperative to analyze each layer's cumulative effect within its respective level of abstraction (whether \textit{strategic}, \textit{logical}, \textit{operational}, or \textit{implementation}).

\begin{enumerate}
\item \textbf{Strategic Layer:} Conceptual, focusing on integrating LLM and GPT capabilities into high-level decision-making processes and predictive analytics tools. At this layer, LLMs and GPTs provide high-level direction and support to the entire 6G architecture. Their predictive analytics capabilities can, for example, forecast network structure and growth, adapt the network to user behavior, and adaptively configure the network, ensuring it remains agile and obeys e.g., regulatory requirements and SLA guarantees.
\item \textbf{Logical Layer:} 
Focused on being integrated into network design tools, this layer ensures a seamless data flow and desired network configuration. Upholding integrity and control over the entire network behavior presents intricate, context-sensitive challenges that continually adapt with user evolution. Here, LLMs can be calibrated to refine data flows, ensuring efficient communication, security, and effective distribution of applications, tasks, and resources. This meets both \enquote{Functional} (driven by users) and \enquote{Non-functional} (driven by architectural quality) objectives. Concurrently, GPTs can be tasked to fine-tune diverse network structures and configurations, perpetually evaluating their security and efficacy. This balances 6G's capabilities between users and their operational demands. Merging these dual capabilities guarantees an ever-efficient, adaptable, and safeguarded network.
\item \textbf{Operational Layer:} This layer is dedicated to overseeing the development and lifecycle of applications, AI, maintenance protocols, security frameworks, and threat detection mechanisms, all while maintaining a real-time monitoring focus. It provides continuous and adaptive support to the real-time capabilities of 6G. Guided by the insights from the Logical layer, LLM and GPT operational frameworks and runtimes can, for example, monitor, control, and prioritize network traffic. They can preemptively pinpoint bottlenecks, define protocols for different users based on their application and service profiles, and adapt contextually to 6G SLA performance standards. This is achieved by channeling resources, decentralizing applications, interfacing with different AI capabilities, dynamically sharing capabilities at different levels of networking, and orchestrating timely reactions to disruptions or emergent requirements.
\item \textbf{Implementation Layer:} Integration into software development and optimization tools, as well as inclusion in hardware analysis and configuration tools. LLMs can be tailored to bolster software enhancements across 6G network elements, ranging from edge devices to assorted network gear, including base stations and Open Radio Access Network (O-RAN) \cite{oran_alliance} modules. The aim here is to ensure a \enquote{continuous seamless integration} of diverse applications and services. Within this layer, the role of LLMs emerges as a vital asset, prepared to be \enquote{trained for purpose}. LLMs can perpetually assess hardware components, guarantee optimal setups rooted in behavioral insights, and ensuring security and compatibility, provisioning reliable and adaptable operations.
\end{enumerate}

The central convergence in the Logical layer highlights the vision of translating the strategic vision into logical steps that can be operationalized. It serves as a bridge, ensuring that strategic objectives are aligned with operational capabilities. This requires a thorough understanding of the service's functional, service, development, operating, and lifecycle requirements and their interrelationships. By focusing on this convergence within the Logical layer, one can ensure that strategic aims are accurately reflected in the system's operational design, leading to a service that is not only strategically aligned but also operationally viable and efficient.

While LLMs can play the role of \enquote{game-changer} in enhancing the AI-native capabilities of 6G systems, it is crucial to recognize and address their inherent limitations to harness their capabilities fully. 

\subsection*{Limitations of LLMs: Need for safeguards and human-in-the-loop}
As LLMs increasingly influence various sectors of modern society, including mobile networks, understanding their limitations becomes fundamental. This section delves into a subset of pressing concerns associated with the deployment of LLMs, including trustworthiness, resource constraints, and the extent of automation. While these represent key areas of focus, it is worth noting that there exist additional nuances and risks that further underline the importance of a balanced approach in utilizing such technologies.

Ensuring the trustworthiness of AI has become a paramount concern in the progression of technology, especially as AI systems are increasingly integrated into critical sectors of society \cite{ding2022roadmap}. Ethical and transparent AI functionalities are essential for societal acceptance and the practical functionality and reliability of AI applications. A leading reference, the European Commission's \enquote{Ethics Guidelines for Trustworthy AI}, emphasizes the significance of ensuring AI systems are lawful, ethical, and robust from both a technical and social perspective \cite{etheu}. Furthermore, the EU's proposed AI Act \cite{com2021laying} aims to establish a risk assessment-based regulatory framework that ensures AI practices in the EU adhere to high safety standards and respect fundamental rights. As AI continues its trajectory of profound influence on global societies and economies, establishing its trustworthiness through rigorous standards and ethical considerations becomes indispensable.

Resource constraints can often hamper the efficiency and capability of deploying ML models. This is particularly true when dealing with large-scale data sets and complex computations, which demand significant memory and processing power. However, advances in hardware acceleration technology, such as graphics processing units (GPUs) and tensor processing units (TPUs), can significantly reduce computational time and allow for more complex modeling \cite{wang2019benchmarking}. Moreover, ML methods, including dimensionality reduction and model optimization, can help mitigate some operational concerns. These techniques can make models more efficient and less resource-intensive, enabling sophisticated computations even on resource-constrained systems \cite{lin2021mcunetv2, lin2022device, brown2020language}.

While automation can significantly improve the efficiency of network operations, it is essential to recognize that there may be inherent limits to the extent to which these processes can be fully automated. Complex tasks often involve variables and considerations beyond the capacity of current AI and ML technologies. Moreover, unforeseen anomalies, exceptions, or crises might demand human judgment and decision-making. This is where the concept of \enquote{human-in-the-loop} solutions comes into play. This approach ensures that while most routine operations can be automated, there remains a human element for oversight, management, and control. The human operator can provide the nuanced understanding, context awareness, and problem-solving abilities necessary to handle complex or unexpected situations. This balance between automation and human intervention can optimize operational efficiency while ensuring the network's robustness and reliability.

Beyond these limitations, it is also essential to touch upon the energy and environmental implications of deploying such advanced models. While this is a critical topic, we will provide an overview rather than an in-depth analysis, as this paper's primary emphasis is on the high-level architectural, management, and orchestrating aspects of LLM-enabled systems.

\subsection*{Energy and Environmental Implications of LLM Deployments in 6G Systems}
The upcoming era of 6G communication systems promises not only a exceptional influence on global growth, productivity, and societal functions but also intersects notably with global sustainability objectives \cite{matinmikko2020white}. The United Nations’ Sustainable Development Goals (UN SDGs) chart a path for a future that seeks to address pressing challenges ranging from poverty alleviation and gender equality to climate change action and urban development \cite{united2022sustainable}. 6G, with its impending commercial launch targeted for 2030, aligns closely with the timeline set for the realization of these global goals \cite{matinmikko2020white}.

The vision of 6G, as proposed in \cite{matinmikko2020white}, looks beyond merely offering communication services. It envisions 6G as a multi-faceted entity: a service provider aligned with UN SDGs, a granular data collection tool for indicator reporting, and a foundational block for future ecosystems that align with these goals. Such ecosystems will harness the capabilities of 6G, targeting goals like smart cities, gender equality, and climate change mitigation. Simultaneously, these advancements in 6G will facilitate breakthroughs in various fields such as virtual learning and smart traveling, contributing further to carbon footprint reduction.

Within this framework, according to \cite{ziegler20206g}, energy efficiency stands as a paramount design criterion for the 6G framework. The network's performance is intrinsically tied to the energy availability across its architectural domains. This focus on energy efficiency is further echoed in the Hexa-X European 6G flagship project, which targets both energy efficiency and the CO2 footprint of network infrastructure as core challenges to be addressed \cite{uusitalo2021hexa}. The importance of this endeavor lies in the fact that ICT technologies, which 6G aims to revolutionize, have a significant carbon footprint on communication networks and wireless terminals. Addressing this will not only reduce the environmental impact but also foster a wider adoption of these technologies in everyday life. In turn, this adoption can lead to optimized operations in sectors like agriculture, transport, and environmental monitoring \cite{uusitalo2021hexa}.

Given the stringent sustainability and efficiency prerequisites of 6G systems, integrating resource-intensive technologies like LLMs demands careful attention. Therefore, it becomes paramount to scrutinize the energy and environmental footprint of the tools steering the 6G advancements. Central to this is the role of LLMs, which intriguingly position themselves as both potential contributors and mitigators within this intricate ecosystem.

In this respect, with growing consciousness regarding the environmental impact of technological advancements and the requisite sustainability of AI \cite{wu2022sustainable}, there has been a noticeable shift towards acknowledging and addressing the resource consumption and carbon footprint intrinsic to the lifecycle management of LLMs. A notable step in this direction has been taken by Meta, who disclosed the electricity consumption and carbon footprint of their LLaMA models \cite{touvron2023llama}. This action aligns with a broader trend, where other technological giants have also unveiled detailed analyses concerning the energy and carbon footprint of prominent models such as Pathways Language Model (PaLM), GPT-3, and Evolved Transformer \cite{chowdhery2022palm, patterson2021carbon, patterson2022carbon}.

Analyzing the figures released for the LLaMA model training process, a massive computational undertaking involved utilizing \textit{two thousand forty eight} 80GB GPUs for an estimated five months \cite{touvron2023llama}. This extensive operation consumed around 2,638,000 KWh of electricity, analogous to the yearly consumption of 1,648 average households in Denmark. The process emitted about 1,015 tonnes of carbon dioxide equivalent (tCO2e), comparable to the annual carbon footprint of 92 Danish citizens.

What sets apart the reporting approach adopted in \cite{touvron2023llama} is its encompassing methodology. Instead of limiting the reporting to the final stages of model training, the entire computational journey, including experimental and unsuccessful runs, has been accounted for, reflecting a comprehensive view of the environmental cost of ML and LLM development. This methodology aligns with the Operational Lifecycle Analysis (OLCA) for ML presented in \cite{dodge2022measuring}. By tracking emissions from the nascent exploratory stages to the deployment of the final model, a holistic understanding of a model's environmental footprint emerges. This view is particularly salient as it captures emissions often overlooked by standard metrics.

     \begin{figure*}[htb!]
    \centering
      \includegraphics[width=0.9\textwidth]{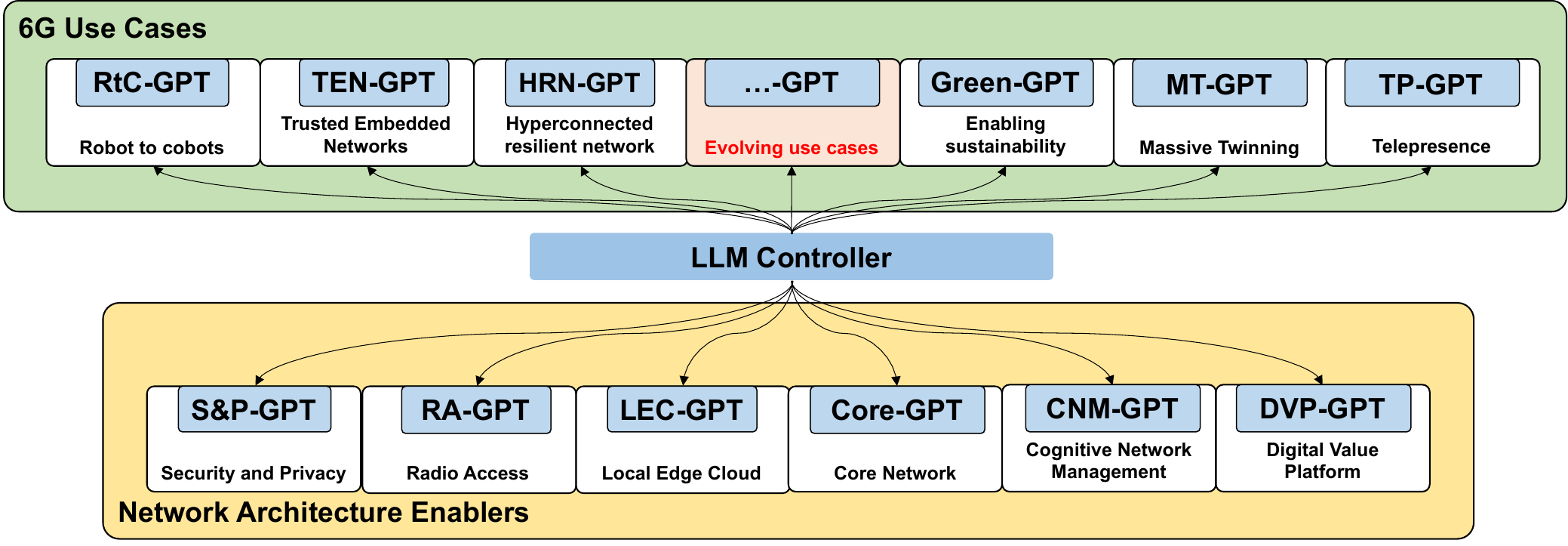}
       \centering
      {\textbf{\caption{Taxonomy of LLMs for 6G, mapping the potential applications of LLMs to both  6G use cases and underlying network architecture enablers. The LLM controller serves as a hub, suggesting the cooperative and management roles LLMs can play in a 6G ecosystem.}\label{fig:3}}}
    \end{figure*}
    
Many optimization strategies geared towards energy-efficient LLMs predominantly focus on refining the model's architecture or pivoting towards sustainable power supplies such as renewable energy sources. While these strategies are vital, given our focus, our interest leans towards solutions intertwined directly with mobile systems architecture at the intersection of 6G and LLMs.

The wave before generative AI saw a distinct trend of training and deploying AI models on energy-efficient platforms such as low-power CPUs, GPUs, or specialized hardware like TPUs. This direction also resonated with efforts to enhance various components of the 5G network \cite{ziegler20206g}. Such hardware specializations not only lead to significant reductions in electricity consumption during the training and inference phases but also contribute to reduced carbon emissions. The development of telco-specific LLMs is likely to benefit immensely from such optimized hardware. Furthermore, much like the emergence of hardware specifically tailored for certain AI tasks, like vision processing units (VPUs) and TPUs for computer vision \cite{sipola2022artificial}, we anticipate the evolution of hardware specialized for LLM execution in the coming years.

From another perspective, the dynamic and heterogeneous nature of mobile networks necessitates a flexible approach to computational resource allocation. This becomes particularly relevant when considering LLMs instances. Guided by real-time workloads and accuracy benchmarks, LLMs should efficiently scale their resource demands to ensure optimized energy usage and minimized carbon footprints. This dynamic resource allocation, coupled with strategies such as caching, memorization, and incremental training, can effectively minimize redundant computations, enhancing overall operational efficiency, also within the 6G landscape \cite{khowaja2023chatgpt}.

Reflecting on all these approaches and emerging directions, it is necessary for the telecommunications sector to embrace comprehensive and transparent methodologies in computing and reporting the environmental footprints associated with the deployment of AI-native network infrastructures, particularly focusing on the integration of LLMs within the 6G landscape.

\section{LLMs and GPTs in 6G: Taxonomy and State-of-the-art}\label{sec:3}
\subsection*{Taxonomy}

LLMs are expected to enhance and support 6G networks in multiple use cases and on multiple sites of its architecture \cite{karapantelakis2023generative, bariah2023large, lin2023pushing}. 
    
Figure \ref{fig:3} illustrates a taxonomy of LLMs mapped to their potential roles in a 6G environment. This taxonomy builds on the fundamental 6G use cases and societal values introduced in \cite{uusitalo20216g}, as well as the business perspectives for 6G described in \cite{yrjola2020white}.

The upper part of the figure underscores the various \textit{6G use cases} LLMs can support. Ranging from \enquote{Robot to cobots} that might involve automating processes and coordination, to the \enquote{Telepresence} that can potentially enhance real-time communication experiences. These use cases touch on hyper-connected resilient networks, sustainability, and other evolving use cases that 6G aims to address. Contrastingly, the lower part of the figure sheds light on the \textit{network architecture enablers} where LLMs might find an application. With components like \enquote{Security and Privacy} emphasizing the importance of safe communications, to \enquote{Cognitive Network Management} which may automate and optimize network configurations, it is evident that LLMs can play a pivotal role in several key network operations.
Central to this taxonomy is the "LLM Controller" It serves as a nexus, connecting both the use cases and the network enablers. This suggests that while there may be models explicitly designed for particular domains (e.g., \enquote{Green-GPT} for sustainability), there is also envisioned a role for a central controller GPT---possibly supported by additional AI components. Such a component would be instrumental in managing these expert models, fostering cooperation among them, and ensuring seamless integration.

Furthermore, these LLMs are not just restricted to high-level operations. We envision their applicability extending to finer-grained models and even being transferred to specific downstream tasks. This includes tasks like beamforming, power management, and handovers, as discussed in \cite{bariah2023large}. As such, the potential for LLMs in a 6G ecosystem is vast, bridging the gap between abstract use cases and the concrete architectural components that enable them.

     \begin{figure*}[htb!]
    \centering
      \includegraphics[width=0.9\textwidth]{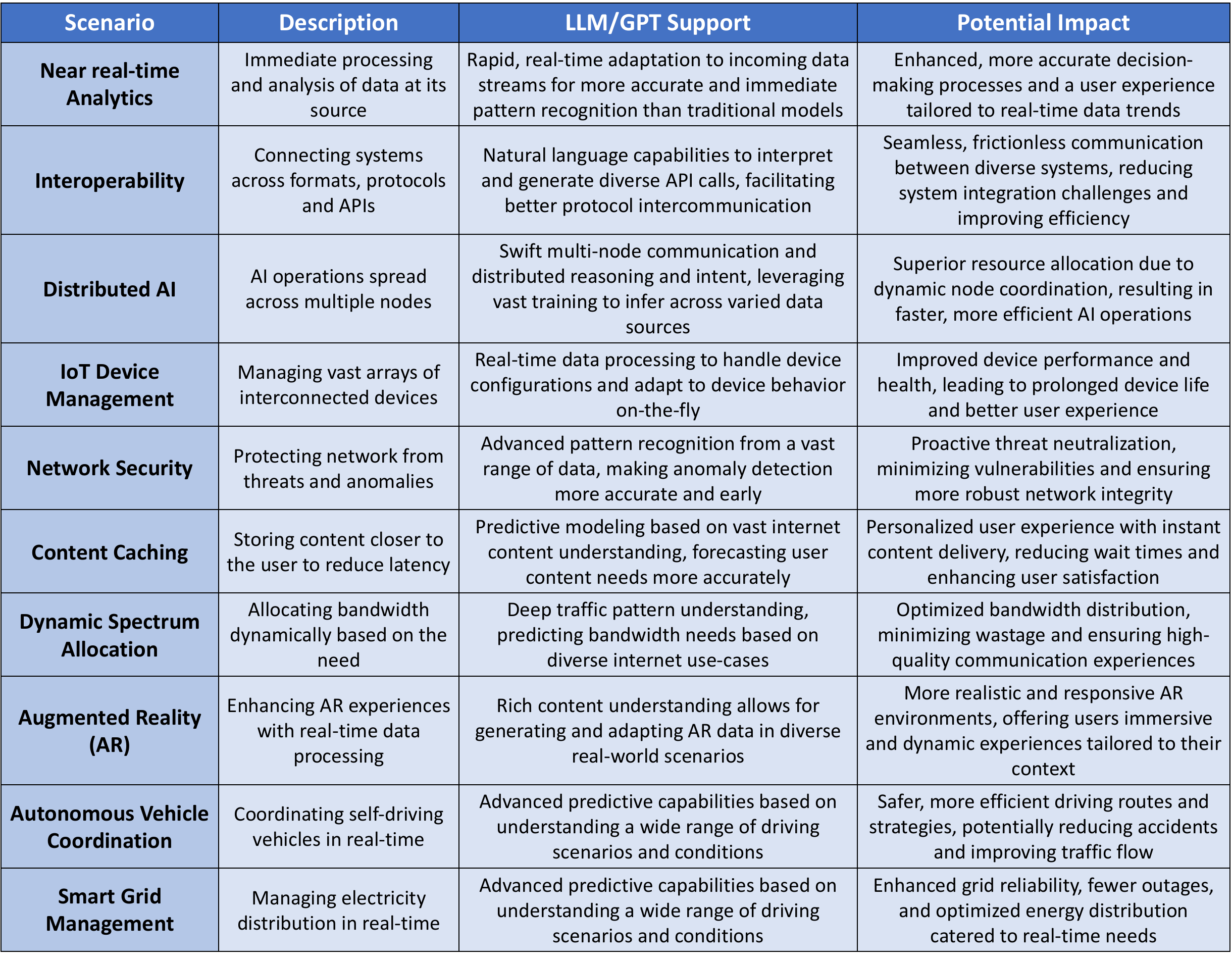}
       \centering
      {\textbf{\caption{Overview of LLMs/GPTs use cases in 6G scenarios. The table highlights the capabilities and potential impacts of leveraging these advanced models, distinguishing them from traditional ML/AI approaches.}\label{fig:table}}}
    \end{figure*}

To exemplify this, consider the increasing possibilities given by Multimodal Large Language Models (MLLMs). Tasked with processing diverse data types, MLLMs exemplify the versatility and adaptability of LLMs \cite{yin2023survey}. These models can help addressing challenges such as data heterogeneity, semantic ambiguity, and signal fading, cementing their role as robust task-solvers \cite{jiang2023large}. Moreover, MLLMs underscore the notion of seamless integration and cooperation among models, as they often function in tandem with unimodal LLMs and other AI components. The duality of their operation, handling both high-level tasks and delving into the intricacies of finer-grained operations, is evident. While traditional LLMs primarily cater to natural language processing (NLP) tasks, MLLMs embrace a wider operational spectrum, enhancing user interaction and communication flexibility with machines \cite{jiang2023large}. Yet, it is important to note that the journey of MLLMs is still in its initial stages. Despite their transformative capabilities, they often deal with the constraint of input-side multimodal understanding, limiting their ability to produce content across diverse modalities. Nevertheless, emerging systems like NExT-GPT aim to redress these limitations, introducing a new set of comprehensive MM-LLM systems suitable for both multimodal understanding and generation \cite{wu2023next}.

While much of the current practical efforts concerning MLLMs are directed towards multimodal signals, including text, audio, image, and video, these are not directly aligned with telco-specific needs. This highlights a pressing need to amplify research and development in this area, as highlighted in \cite{bariah2023large}. Nonetheless, Figure 4 provides a glimpse into how such integration could potentially operate within the telecommunications context, illustrating potential applications of how MLLMs might process multimodal data \cite{lahat2015multimodal}.

     \begin{figure*}[htb]
    \centering
      \includegraphics[width=0.8\textwidth]{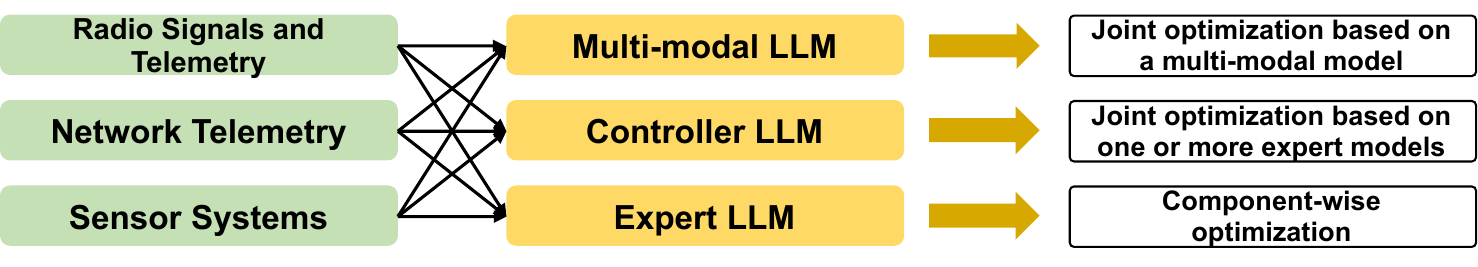}
       \centering
      {\textbf{\caption{Illustration of three LLM processing approaches for multimodal data. The scenarios include inputs from radio signals and telemetry, network telemetry, and sensor systems. The depicted approaches are: a multimodal capable LLM, a controller LLM for processing and result aggregation, and component-wise LLMs without aggregation.}\label{fig:4}}}
    \end{figure*}

While Figure \ref{fig:3} offers a conceptual overview of the interplay between LLMs and the broader 6G ecosystem, a more granular perspective is necessary to fully appreciate the potential of these models in this space. The subsequent table, presented in Figure \ref{fig:table}, elucidates specific scenarios where LLMs/GPTs can be particularly transformative compared to the current state-of-the-art. In fact, while traditional ML/AI methodologies have made significant strides in these domains \cite{letaief2019roadmap, peltonen20206g}, LLMs/GPTs, given their vast training data and swift adaptability, can potentially offer more nuanced solutions. These models inherently possess the capability to quickly discern patterns from massive datasets and adapt to new information at an unprecedented scale \cite{tamkin2021understanding}. This makes them particularly suited for dynamic scenarios, like those associated with 6G, where real-time adaptability and extensive knowledge bases are paramount.

For instance, in scenarios like 'Near real-time Analytics' and 'Network Security', the advantage is not merely about processing data quickly — it is about the depth of insight and foresight that models like GPTs can provide because of their comprehensive training. The table delineates how LLMs/GPTs can be leveraged across these 6G scenarios, underlining their potential transformative impacts.

However, it is crucial to understand that this table touches upon just a fraction of the myriad applications LLMs/GPTs can assist with in a 6G environment. As we advance further into this technological era, it is likely that the unique capabilities of LLMs/GPTs will find relevance in even more areas (as highlighted in \cite{karapantelakis2023generative}), some of which we might not have even envisioned yet.

Building on this, it is essential to categorize the different types of LLM/GPT models available, as they can vary based on their scope, application, and accessibility. We delineate four principal LLM/GPT categories, with models potentially being open source or reliant on closed APIs. LLMs are anticipated to be accessible via standardized embedding formats and open APIs, such as those offered by OpenAI \cite{openai} or LangChain \cite{langchain}. The four categories are:

\begin{enumerate}
  \item \textbf{Foundation:} A universal LLM model that is not tailored to a specific domain. Although foundational, this model can be enhanced through prompt programming and fine-tuned for specific use cases.
  \item \textbf{Specialized:} An LLM model tailored for a particular user or application, available in either closed or open-source fashion. Examples of its use include chat applications and instruction generation tools. It can also be adapted to environments with limited resources.
  \item \textbf{Hybrid:} This LLM model combines the broad, general knowledge of the foundation model with the specialized expertise of the specialized model.
  \item \textbf{Controller:} Driven by LLM-based autonomous agents, this model functions as a controller, linking other models and system components. It operates either autonomously or under human oversight.
\end{enumerate}

\subsection*{State-of-the-art towards the AI Interconnect}

As previously discussed, LLMs are anticipated to advance 6G networks across several use cases and various facets of its architecture. Due to its prominence as a rapidly evolving research area, there have been a surge of studies aiming to delineate how LLMs might be integrated within the mobile network context. These studies span from addressing specific challenges with pinpointed solutions to casting a wider net by offering a panoramic view of the vision, challenges, and opportunities intrinsic to this new domain. 

\begin{figure*}[htb]
    \centering
      \includegraphics[width=1\textwidth]{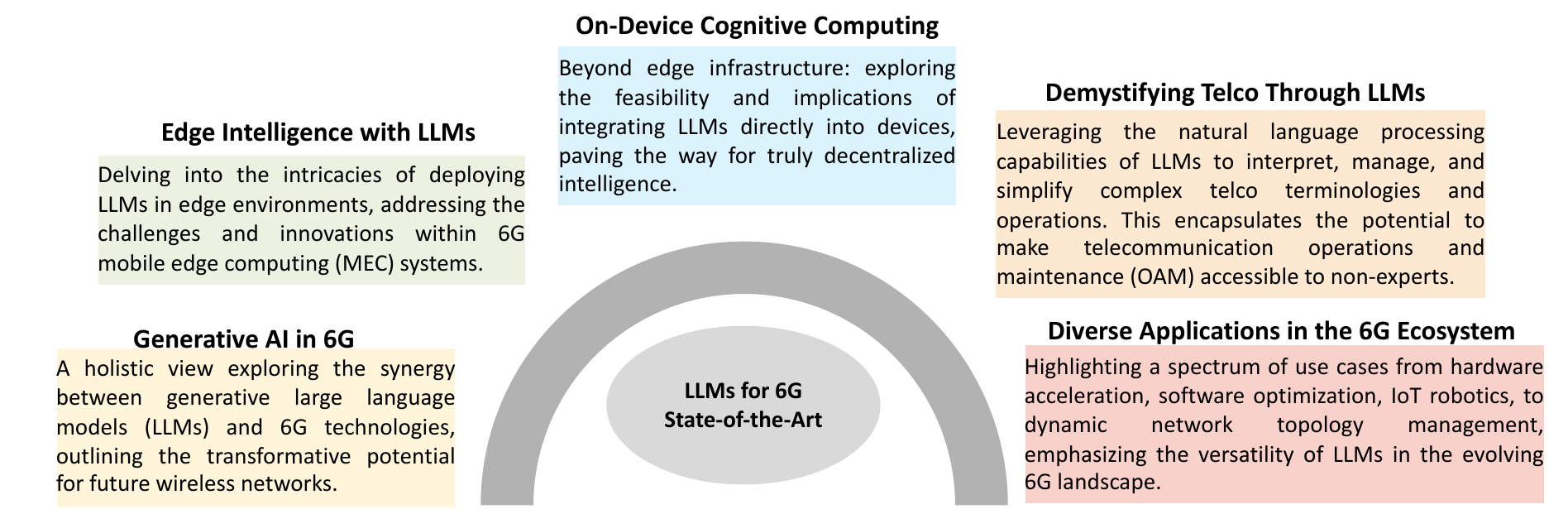}
       \centering
      {\textbf{\caption{Visual representation of the categorization of LLMs state-of-the-art in the 6G context.}\label{fig:sota}}}
    \end{figure*}

Figure \ref{fig:sota} provides a visual representation of these studies categorization, charting a path from the core concepts to the broader applications. As we progress through this section, we will delve deeper into each layer of the figure, introducing the complexities and insights of the associated research.

\subsubsection*{Generative AI in 6G networks}

The potential of GenAI in revolutionizing mobile telecommunications networks has been a focal point of recent research. A comprehensive review presented in \cite{karapantelakis2023generative} offers insights into the developments, challenges, and classification approaches surrounding GenAI in this domain. Besides a historical perspective on the evolution of GenAI, the study underscores its growing significance over LLMs within mobile telecommunications. It further highlights existing challenges and potential areas warranting future exploration.

In a similar fashion, the work in \cite{bariah2023large} emphasizes the transformative potential of LLMs in reshaping wireless networks. By articulating a vision where LLMs initiate a new era of autonomous wireless networks, the study expounds on two central themes: 'LLMs for Wireless' and 'Wireless for LLMs.' The former elucidates the multifaceted capabilities of LLMs, encompassing tasks such as beamforming, channel estimation, and modulation, and bridging radio signals with other modalities. The latter envisions an interconnected 6G network ecosystem anchored by devices with embedded LLMs, shifting the paradigm from data-centric to knowledge-centric networks.

Building on the edge-cloud paradigm, the NetGPT proposal presented in \cite{chen2023netgpt} charts a pathway to strategically deploy LLMs at both edge and cloud, capitalizing on their unique computational attributes. Edge-based LLMs promise lower latency and enhanced privacy, drawing from local information for interactions with their cloud-based counterparts. The practicality of NetGPT is showcased by deploying open-source models like GPT-2-base \cite{radford2019language} and LLaMA \cite{touvron2023llama}, harnessing a lightweight fine-tuning method rooted in low-rank adaptation. Integral to this architecture is the establishment of a control plane in the RAN for NetGPT, underpinning the synchronization of prompting and generative information transmission. This holistic approach signifies strides towards a unified solution that amplifies network optimization through intelligent edge-deployed LLMs.

In the exploration presented in \cite{maatouk2023large}, the study sheds light on the rapidly evolving landscape of LLMs within the telecom industry. The authors demystify the capabilities and limitations of current LLMs, highlighting their strengths in semantics and generative abilities. Focusing on tangible use cases like network anomalies resolution, 3rd Generation Partnership Project (3GPP) specifications comprehension, and network modeling, they demonstrate the immediate value LLMs can bring to telecom. The work not only emphasizes the existing limitations such as hallucinations, fabrication, and computational complexities but also accentuates forthcoming research avenues, including foundation models in telecom, LLM compression, privacy considerations, and sustainability impacts.

\subsubsection*{Edge Intelligence with LLMs}

As 5G continues its evolution towards 6G, the emphasis on edge-centric designs remains undiminished. With the relentless push to extend AI capabilities like LLMs to the edge, the dynamics of this transition become particularly intriguing. However, given the peculiar requirements of LLMs, migrating such advanced capabilities to edge intelligence is considerably complex. Numerous studies have begun to delve into this context, exploring the potentialities and challenges that arise when integrating LLMs into the edge-driven architecture of emerging networks.

A contribution in this area introduces an autonomous edge AI framework. Anchored by a hierarchical cloud-edge-client blueprint, LLMs like GPT are poised in the cloud, facilitating language understanding, planning, and code generation, while specialized AI models are situated at the edge. This design leans towards a user-centric approach, dynamically scripting necessary codes for new model training via edge federated learning \cite{shen2023large}.

The work presented in \cite{lin2023pushing} offers a deep dive into the transformative potential of LLMs when merged with 6G mobile edge computing (MEC) systems. While LLMs offer unmatched capabilities, their cloud deployment poses challenges, including latency, bandwidth costs, and privacy. As 6G moves towards standardization, integrating LLMs at the edge becomes imperative. The study underscores the feasibility of this integration by advocating advanced AI methodologies. Techniques like split ML, parameter-efficient fine-tuning, and model quantization emerge as vital, promising significant reductions in model training times without sacrificing performance. By spotlighting applications in healthcare and robotics, the research amplifies the urgency of edge-based LLMs. Yet, challenges like communication demands, computational intensities, and storage concerns loom large. To navigate these, the authors propose a robust 6G MEC structure optimized for LLM deployment, made up of modules for network management, model caching, and training.

The synergy of LLMs, edge networks, and multi-agent systems for enhanced wireless networks is insightfully explored in \cite{zou2023wireless}. Contrasting cloud-based LLM limitations, the study introduces on-device multi-agent LLMs for self-directed, edge-situated intelligent decision-making. A game theoretic approach emerges for these multi-agent LLMs, leading to applications like intent-driven network automation. A featured case study delves into energy savings and user transmission rates, underscoring the transformative potential of this technology. In essence, this work paves a novel path for wireless network design, emphasizing the fusion of collective intelligence and GenAI capabilities.

\subsubsection*{On-Device Cognitive Computing}

On-Device LLMs execution has recently been proposed for multi-agent GenAI, in which generative agents collaborate to solve network intents \cite{zou2023wireless}. This study delves into the fusion of LLMs, edge networks, and multi-agent systems for advanced wireless networks. This convergence aims for self-governed, edge-located intelligent decision-making. The work introduces on-device multi-agent LLMs that collaboratively tackle network tasks. It contrasts this with the constraints of cloud-based LLMs and provides insights into a game theoretic approach for multi-agent LLMs. Key applications for this technology, such as intent-driven network automation, are discussed along with a case study on energy savings and user transmission rates. This work sets a new direction for wireless network design, emphasizing collective intelligence and GenAI potentials.

As the field rapidly progresses, there is an ever increasing research initiatives focusing on optimizing LLM execution in constrained devices. The challenge lies in achieving this without significantly decreasing the quality of inference. Several recent studies have championed innovative techniques in this regard, pointing towards a promising future where advanced AI capabilities can be seamlessly harnessed in resource-limited settings \cite{lin2023awq, xiao2023smoothquant, tinychat}.

\subsubsection*{Demystifying Telco through LLMs}
The possibility of integrating LLMs into the telecommunication sector has been gaining momentum, also driven by the potential to make complex technical procedures more accessible. A relevant initiative in this context is the fine-tuning of LLMs for 3GPP terminology \cite{bariah2023understanding} and the broader telecom domain \cite{ericsson2022neural}. These specialized LLMs can scrutinize technical documentation, respond to queries, and even aid in network debugging and deployment.

Similarly, the work introduced in \cite{soman2023observations} stands as an experimental exploration into consolidate LLMs with conversational interfaces tailored for the telecommunications sphere. Their thorough analysis investigates the capability of these models to adapt to domain-specific jargon, especially within the telecom context, while also assessing their contextual retention across diverse interactions.

Large software-driven firms, still operating in the telco industry, face challenges in promptly pinpointing duplicate trouble reports (TRs) \cite{bosch2022fine}. A BERT-inspired \cite{devlin2018bert} system was conceived to tackle this, yet faced constraints with TRs outside its training domain. Integrating domain-specific telecommunication knowledge emerged as a remedy, bolstering both the system's efficacy and adaptability \cite{bosch2022fine}.

In order to enhance the telecommunication aptitude of LLMs, particularly when conventional models like GPT-3 \cite{brown2020language} stumble due to data scarcity in niche areas, the Telecom Knowledge Governance (TKG) initiative was presented in \cite{cai2023tkg}. By refining these models using a rich telecom-specific corpus, the initiative promises more accurate, domain-centric responses.

The work presented in \cite{maatouk2023teleqna} introduces \enquote{TeleQnA}, a benchmark dataset designed specifically to gauge the proficiency of LLMs in the telecommunications sector. Comprising 10,000 \textit{Question and Answer (QnA)}  sourced from various documents, including research articles and standards, the dataset \cite{teleqna} is a product of an automated QnA generation framework, enhanced with human oversight. Evaluations using this dataset reveal that while LLMs, such as GPT-3.5 and GPT-4, can competently address broad telecom queries, they falter with intricate standards-based questions. Incorporating telecom-specific context notably boosts their accuracy. Interestingly, when compared to active telecom professionals, LLMs exhibit competitive performance, emphasizing their transformative potential in the field.

Lastly, initiatives like the ones shown in \cite{bubbleran} stand as proof of the transformative nature of LLMs in the telecom sector. By translating complex technical steps into simple conversational directives, they empower even the 'non-expert' to supervise and roll out networks, showcasing the democratizing potential of LLMs in advanced network administration.

\subsubsection*{Diverse Applications in the 6G Ecosystem}

The versatility of LLMs offers a multitude of opportunities that can profoundly influence the 6G ecosystem across a myriad of applications, spanning hardware acceleration, software optimization, and dynamic network topology management. When delving into specific scenarios and applications, the possibilities seem boundless. The IoT and robotics example provided below is merely a representative example, underscoring the extensive potential of LLM applications.

The work presented in \cite{vemprala2023chatgpt} underscores the transformative potential of GPT principles in robotics. Instead of the traditional, labor-intensive manual coding for robot tasks, GPTs enable a more intuitive high-level feedback mechanism, allowing even non-experts to oversee wireless or wired robot functionalities. Such an approach revolutionizes robotics, facilitating the generation of diverse task codes from simple puzzles to intricate operations in manipulation, aerial, and navigation domains.

Addressing the limitations of conventional deep offloading in mobile edge computing, the research in \cite{dong2023lambo} introduces the LAMBO framework. With its integration of LLMs, LAMBO encapsulates four vital components to improve data representation, decision-making, pre-training, and adaptability in dynamic settings. By offering solutions to the challenges in MEC systems, LAMBO stands as a promising tool for future edge intelligence, backed by promising simulation results.

A fascinating intersection of LLMs and wireless networking technologies is showcased in \cite{du2023power}. This research highlights LLMs' capabilities in generating hardware description language (HDL) code. Focusing on the intricate wireless communication domain, the paper reveals how LLMs can boost productivity in code refactoring, reuse, and validation. Particularly, the study champions the use of LLMs in crafting HDL code for wireless systems, emphasizing its potential to reshape the landscape of wireless signal processing.

Lastly, the challenges in analyzing network topologies and communication graphs are addressed in \cite{mani2023enhancing}. Here, LLMs are harnessed to transform natural language questions into task-specific code, providing a more intuitive network management experience. The study introduces a novel framework that leverages the latest advancements in LLMs for crafting code tailored for graph management. While the prototype demonstrates competence in simpler tasks, there is room for improvement in complex operations. The research sets the stage for further exploration, emphasizing the need to refine domain-specific techniques and ensure code correctness.

In essence, the growing momentum in research related to LLM applications within the 6G framework underlines the boundless opportunities awaiting. This wide spectrum of use-cases signals a promising future where advanced AI capabilities are seamlessly integrated into the very fabric of the 6G ecosystem.

\section{Architectural Foundations for Intelligent Evolution: From 5G to AI Interconnect}\label{sec:4}

The evolution of 5G since its first release has not only been characterized by enhanced connectivity and performance but also by a strategic alignment towards a more flexible, responsive, and intelligent architecture spanning from the RAN to the Core network. This journey included multiple advancements ranging from evolved network functions, service provisioning and orchestrated operational patterns, to the integration of AI capabilities.

However, while 5G has successfully fostered a favorable environment for the deployment and functionality of AI applications, its native architecture cannot be considered as \enquote{AI-native} \cite{li2017intelligent}. In fact, while 5G architecture is proficient in provisioning AI applications, it does not inherently possess built-in AI processing or decision-making faculties. Consequently, essential AI-driven tasks such as real-time data analysis and predictive maintenance are primarily handled by supporting AI systems and functionality, introducing additional complexities and potential latency in network operations. Furthermore, 5G has embraced enhanced operational agility through the definition of diverse communication patterns and a service-based architecture, fostering a flexible and scalable network ecosystem. These elements, along with the exploration of intelligence integration, comprise the core of this section, depicting a holistic picture of the 5G evolution and its trajectory towards a more intelligent and adaptable network infrastructure.
\newline
\newline
In the 5G system architecture \cite{5GSYSTEM}, standardized by the 3GPP, a service bus plays a crucial role in interconnecting various network functions. This service bus employs two primary communication patterns: request-response and subscribe-notify \cite{5GSYSTEM}. The request-response pattern is a synchronous method that requires immediate responses, such as setting up a user's connection or modifying a session. This pattern allows for direct communication between network functions, facilitating efficient data exchange. Contrarily, the subscribe-notify pattern provides an asynchronous notification capability, which is particularly useful for reacting to network or application-related events. In this pattern, network functions can subscribe to specific topics and receive notifications when other functions publish messages to these topics \cite{29591}. 
The decoupling of network functions, enabled by the Service-Based Architecture (SBA) \cite{sba3gpp}, has contributed to the increased flexibility of the 5G architecture. In fact, it enables the network functions to be provisioned and scaled independently, fostering adaptability and resilience in the network, crucial for efficiently supporting a plethora of services and applications \cite{zeydan2022service}. 

5G orchestration also emerges as pivotal aspect of the 5G network architecture, being responsible for the automated management of network services and resources \cite{5GSYSTEM}. It orchestrates various network functions and services, such as setting up network slices, managing resources, and ensuring Quality of Service (QoS) for diverse applications \cite{taleb2017multi}. Network slicing--a significant feature of 5G--enable the set up of multiple virtual networks on a single physical infrastructure. Each slice can be tailored to meet the specific needs of a particular service or application, and the orchestration layer manages these slices \cite{afolabi2018network}. Additionally, 5G orchestration handles the real-time allocation of network resources (e.g., bandwidth and computational power), adjusting these allocations as needed to optimize network performance. It also ensures the QoS for different applications and services by prioritizing network traffic based on factors like the type of application, the user's SLA, and current network conditions. 

Within this orchestrated ecosystem, the 3GPP has devised key 5G Core components such as the Network Data Analytics Function (NWDAF) and the Management Data Analytics Function (MDAF) to bolster the network's intelligence of 5G networks \cite{garcia2023network, ghosh20195g}. The NWDAF collects and analyzes network-wide information, enabling an intelligent, real-time understanding of network conditions and performance. This aids in making predictive decisions to enhance overall network performance. The MDAF serves a complementary role by focusing on collecting and analyzing management data. This includes performance metrics, fault information, and configuration data from network entities, allowing for a holistic view of the network status and facilitating data-driven network management decisions. Both NWDAF and MDAF functions are central to the 3GPP's approach to applying AI and ML techniques for more efficient, responsive, and adaptable 5G networks.

Looking at the RAN, there has been a surge in R\&D efforts aimed at fostering AI-driven radio networks in recent years \cite{yao2019artificial, lee20206g, brik2023survey, mudvari2021ml}. In line with this trend, the O-RAN Alliance \cite{oran_alliance} emerges as a central initiative, committed to revolutionizing the RAN through open standards and architectures. By advocating for open interfaces, disaggregated network components, and a heightened focus on AI-driven intelligence and automation, the O-RAN Alliance aspires to cultivate a more innovative, interoperable, and economically efficient wireless network ecosystem \cite{singh2020evolution, O-RAN}.

     \begin{figure*}[ht!]
    \centering
      \includegraphics[width=1\textwidth]{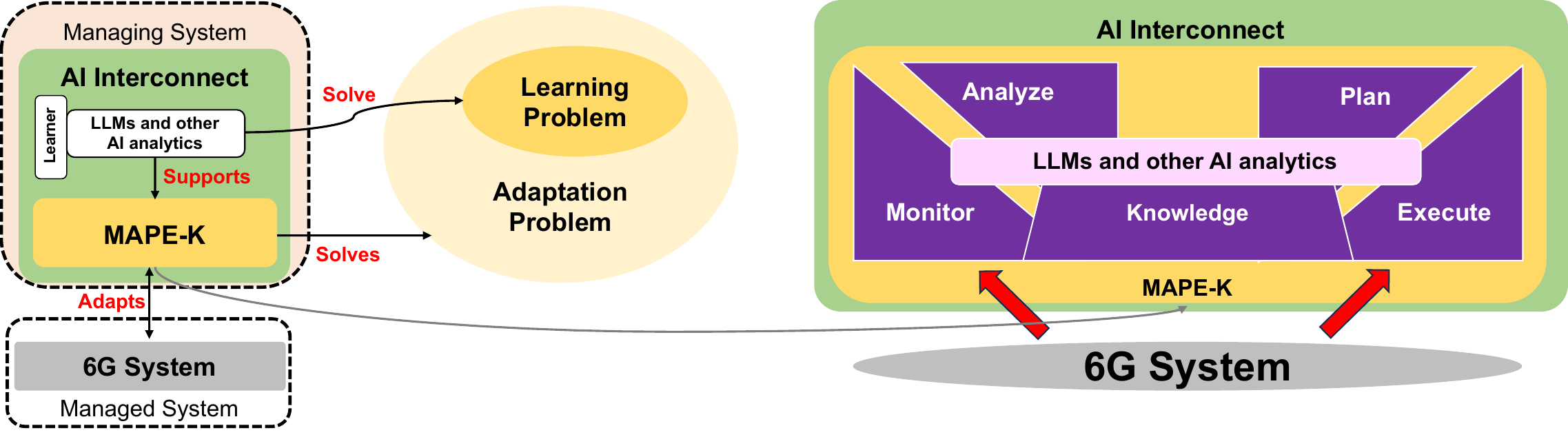}
       \centering
      {\textbf{\caption{MAPE-K as enabling paradigm of the AI Interconnect. On the left, relation of learning problem and adaptation problem with the components of the managing system. On the right, structure of an autonomic element. Elements interact with other elements and with human programmers via their autonomic managers.}\label{fig:mape}}}
    \end{figure*}
    
O-RAN operates based on two Radio Intelligent Controller (RIC) designs \cite{polese2023understanding}. The Non-real-time RIC is configured to manage operations that last several seconds, making it ideal for AI/ML training and aiding service provision. Conversely, the Near-real-time RIC is optimized for operations that range from tens of milliseconds to a second, aligning well with the management of RAN control primitives and the execution of inference tasks. The Near-RT RIC facilitates the deployment of xApps, applications interfacing with the RAN elements to perform specific control and optimization functions. The O-RAN E2 interface serves as the main channel for essential data for RICs, learning, and inference processes \cite{O-RAN-AI}.
O-RAN incorporates an AI/ML framework that significantly transforms the RAN architecture, infusing it with intelligent functionalities \cite{O-RAN, O-RAN-AI}. This framework is anchored on the RIC, housing the xApps—applications employing AI/ML algorithms to automate and optimize various RAN functionalities. The RIC, interfacing with both non-real-time (Near-RT RIC) and near-real-time RAN domains, enables a dynamic information exchange, promoting a more adaptable and responsive RAN. It uses standardized, open interfaces to foster interoperability, minimize vendor lock-in, and encourage a diverse ecosystem of innovative RAN solutions.
The internal messaging infrastructure of O-RAN connects xApps, platform services, and interface endpoints. While there’s no specific technology prescribed (for instance, the O-RAN Software Community has delineated the RIC Message Router or RMR \cite{O-RAN-RMR}), the system is required to meet certain standards. It should enable registration, discovery, and removal of endpoints, like RIC components and xApps, and should provide APIs for direct messaging or through pub/sub methods, guaranteeing efficient routing and data protection \cite{19502804fcb7443fa4dfdc54b802471d}.

The outlined architectural advancements in 5G and beyond, particularly the introduction of analytics-based orchestration and open, intelligent RAN concepts, depict a trajectory towards more integrated, flexible, and intelligent network architectures. These improvements form a foundational basis that seems conducive for the native, seamless incorporation and functioning of advanced AI technologies such as LLMs and GPTs in the upcoming 6G networks.

\section{The 6G AI Interconnect Framework}\label{sec:5}

The envisaged AI Interconnect framework for 6G leverages the power of advanced AI/ML models, such as LLMs and GPTs, coupled with enhanced data analytics capabilities, to comprehend and interpret the vast volumes of data traversing the network. This analysis allows the AI components to discern patterns and trends, predict network congestion, and make insightful decisions about routing data and component placement to optimize the use of radio and network resources. Moreover, the AI Interconnect can diagnose, analyze, and manage network and application tasks in real-time across the edge-cloud continuum.
Central to this approach is the MAPE-K feedback loop, a fundamental architectural model in the design of self-adaptive systems and autonomous computing \cite{gheibi2021applying}.
    
The MAPE-K model outlines a cyclical process enabling systems to self-manage and adapt to evolving conditions for optimized performance. It encompasses four interdependent stages: \emph{(i)} Monitoring, where the system's environment and performance are observed; \emph{(ii)} Analysis, where data gathered is evaluated to understand the system's status; \emph{(iii)} Planning, where strategies are formed based on analysis results; and \emph{(iv)} Execution, where these strategies are implemented. The shared 'Knowledge' component supports all these stages, providing a centralized information base.

In the context of 6G networks (Figure \ref{fig:mape}), the AI Interconnect functions as the \enquote{Managing System}, deploying the MAPE-K loop as a strategic tool to facilitate intelligent control and adaptability in the \enquote{Managed System}. Specifically aimed at optimizing quality properties essential for 6G operations, the AI Interconnect uses the MAPE-K loop to address complex network challenges. LLMs and GPT technologies, within this setup, are instrumental in enhancing the functionality of the MAPE-K loop. They are particularly significant in addressing learning problems within the broader adaptation challenges, enriching the AI Interconnect's capability to execute timely and informed adjustments in the 6G network's operations. The sophisticated processing capabilities of LLMs allow them to manage extensive datasets efficiently, offering critical insights that enhance the \textit{Analysis} and \textit{Planning} stages of the MAPE-K loop, subsequently fine-tuning the adaptive responses of the 6G network.

The AI Interconnect aims to empower the MAPE-K stages to leverage LLMs, and even undertake MAPE-K iterations through a singular LLM or a set of LLMs. This AI-driven interconnect approach offers several significant advantages over traditional network interconnects. Primarily, it grants the network the dynamism required to adapt to evolving conditions, such as fluctuations in data traffic or modifications in network topology. Consequently, the network can sustain optimal performance even when navigating through challenging operational terrains. Moreover, the synergy between AI interconnect and orchestration emerges as a crucial catalyst in realizing Intent-Based Networking (IBN) \cite{mcnamara2023nlp}. IBNs have the capability to translate users' business objectives into strategies for network configuration, operation, and maintenance \cite{leivadeas2022survey}. The integrated use of MAPE-K, AI Interconnect, and GPT/LLM technologies orchestrates a robust, forward-thinking paradigm for managing and optimizing 6G networks. Collectively, these elements create a dynamic, self-adapting network infrastructure, capable of effectively meeting the demands of future connectivity with unparalleled efficiency and intelligence.

     \begin{figure*}[ht!]
    \centering
      \includegraphics[width=0.8\textwidth]{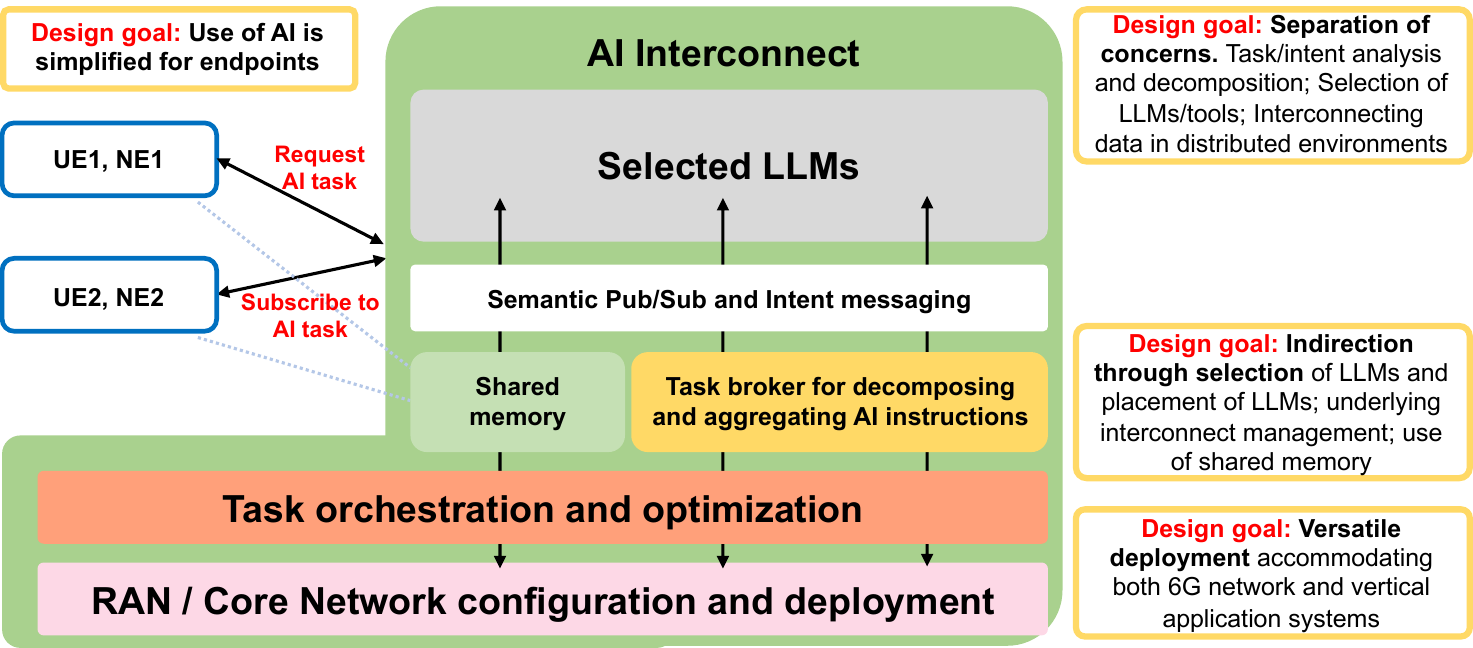}
       \centering
      {\textbf{\caption{GPT-Driven AI Interconnect Architecture. Our approach is structured around four pivotal design goals, from simplifying end-user interactions to ensuring versatile deployment across varied contexts, all rooted in foundational components for optimum efficiency and adaptability.}\label{fig:5}}}
    \end{figure*}

Figure \ref{fig:5} delineates our envisioned AI Interconnect system architecture in detail. Building upon preceding concepts of on-device and in-network LLMs, our blueprint proposes an edge-cloud continuum orchestration, allowing LLMs to be strategically positioned across various network elements, ranging from the user equipment (UE) to the public cloud. 
In developing our system, we contemplate a foundational architecture driven by four design goals and three fundamental components. The first design goal, \textit{simplicity at the endpoints}, advocates for a redistribution of complexity away from the network edge, resulting in a simplified end-user interface and device requirements. The second goal, \textit{separation of concerns}, ensures that different system functionalities are divided into separate components, each having a distinct responsibility, thereby fostering modularity and scalability. Our architectural vision also embraces \textit{indirection through selection} as third design goal. This introduces an intermediary layer optimized for system interactions, fostering a robust adaptability and flexibility in the face of varied operational conditions. \textit{Versatile deployment} rounds off our design goals, emphasizing the architecture's inherent capability to proficiently navigate both network-level and application-level concerns, ensuring a comprehensive functionality across a spectrum of contexts and use cases.

The realization of these design goals is reflected in the mobile network, through the three components depicted in Figure \ref{fig:1}: \emph{(i)} the \textit{control plane}, which is responsible for the orchestration and coordination of network-centric activities such as resource allocation and network configuration; \emph{(ii)} the \textit{user plane}, which handles the data transmission, ensuring the efficient and reliable management of user data packets; \emph{(iii)} the higher-level \textit{application logic}, which servers as the business layer that supports end-user services, analytics, and advanced applications.
All these layers are meant to utilize 3GPP and O-RAN interfaces, promoting a more open and intelligent network architecture. Within this framework, the RIC, complemented by the insights derived from NWDAF and MDAF, plays a central role in orchestrating and controlling these interfaces, acting as the system's cognitive core. This enhanced  \enquote{brain} navigates informed decisions about resource allocation, service provisioning management and orchestration, signal processing, and optimization pathways for network efficiency. This comprehensive design approach can ensure that our system is robust, versatile, and ready to meet the demands of future AI-native network infrastructure.

We base the AI Interconnect and orchestration design on three fundamental concepts:

\begin{itemize}
    \item \textbf{Semantic Publish/Subscribe and Intent-based Messaging.} Our AI Interconnect is designed with a specialized focus on multi-layer semantic request/reply and pub/sub mechanisms. The interconnect is multi-layered to meet the diverse requirements of AI applications, such as the delivery of raw data, prompts and prompt fragments, inference results, and model updates. This approach ensures that endpoints are decoupled across spatial, temporal, and synchronization dimensions. At the core of this design lies a message-based interconnect system that carries more than just data; it encompasses essential AI-related metadata, providing a detailed account of AI operations and forming the basis for essential oversight mechanisms. This inherent attribute promotes accountability by keeping a detailed record of these operations, enabling a rigorous audit of AI-driven processes.
    \item \textbf{LLMs as Controllers.} In our architecture, LLMs are re-imagined as central orchestrators, strengthened with the expertise of specialized models and systems. They bring an element of dynamic flow control, introducing a layer of agility and adaptability to the system. Positioning LLMs centrally as controllers equips the system with the capability to expertly navigate and supervise complex AI landscapes.
    \item \textbf{LLMs as Dynamic Tool Builders and API Brokers.} Beyond their role as controllers, LLMs are also envisioned as dynamic tool creators. They have the capability to dynamically craft tools that are tailored to specific operational needs, ensuring a high degree of flexibility and precision in AI tasks. Furthermore, LLMs act as API brokers, facilitating the seamless interaction and integration between diverse AI tools and platforms, simplifying operations and enhancing the coherence of the AI ecosystem.
\end{itemize}

Another aspect to explore in the context of LLMs is the potential for semantic compression and communication \cite{guo2023semantic,jiang2023large, park2023towards, nam2023language}. This innovative capability can lead to additional possibilities for efficient data transmission and interaction, paving the way for more advanced and intelligent communication paradigms within the AI Interconnect. It could potentially enhance the roles of LLMs as controllers, tool builders, and API brokers by introducing a new layer of semantic intelligence to these functions. However, a deep dive into this aspect of LLM capabilities lies beyond the scope of this paper, but it signifies an intriguing direction for future research and exploration.

Our AI Interconnect seamlessly integrates with open edge-cloud continuum APIs and execution environments, leveraging existing standards and platforms, such as O-RAN, to ensure a unified, interoperable execution framework that gracefully spans across the edge and the cloud.

\section{Semantic Publish/Subscribe and Intent-based Messaging}\label{sec:6}

In the complex landscape of LLMs, there is not a universally recognized \enquote{stack} yet. However, for the purpose of our AI Interconnect, we identify several crucial components, or \enquote{layers} that play a vital role in the functioning of LLMs within this framework:

     \begin{figure*}[htb]
    \centering
      \includegraphics[width=0.7\textwidth]{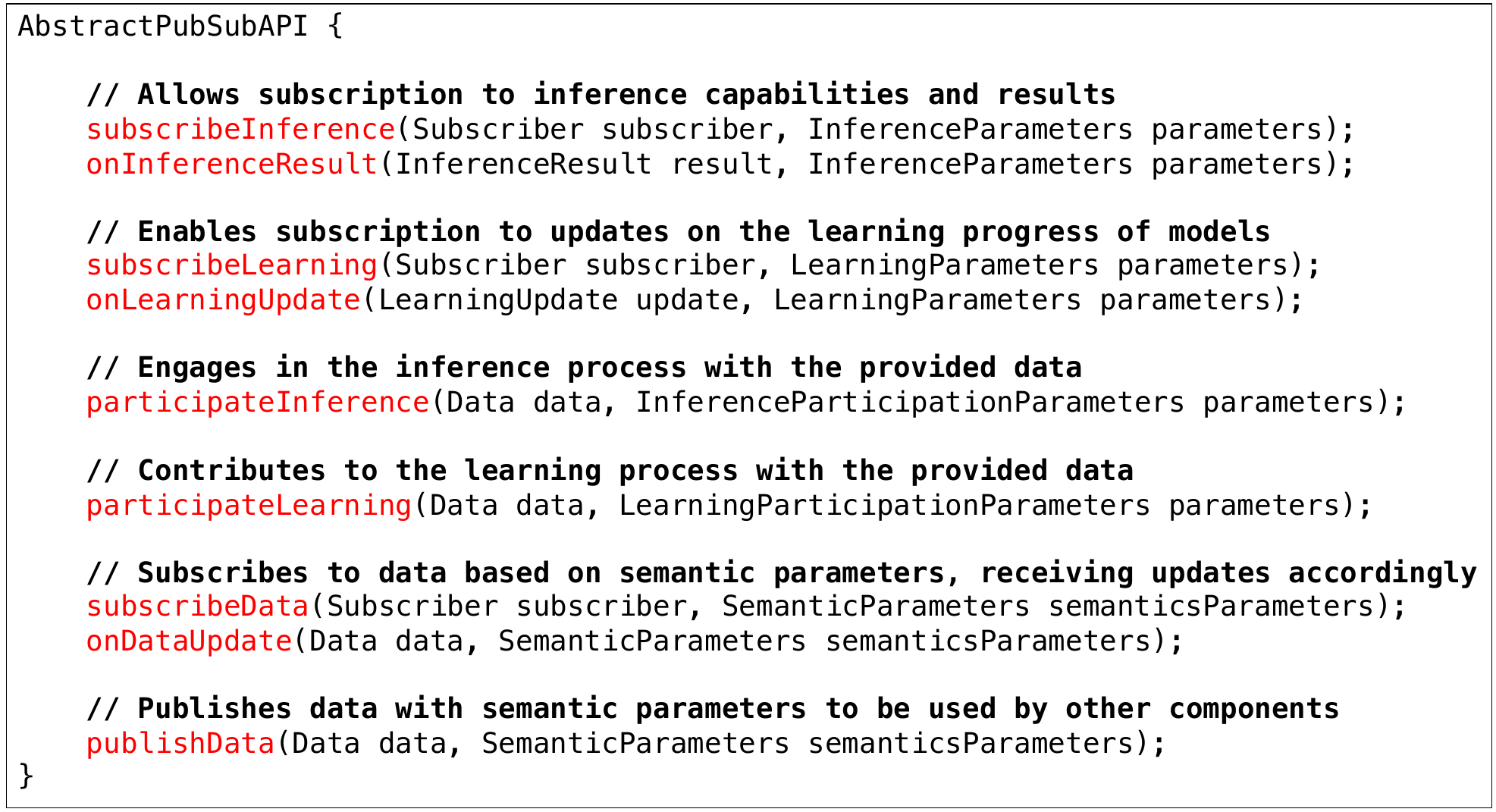}
       \centering
      {\textbf{\caption{Pseudo-code representation of the PubSubAPI interface detailing the methods for subscribing to and participating in inference and learning processes, as well as for managing data flow based on semantic parameters within the AI Interconnect system.}\label{fig:6}}}
    \end{figure*}
    
\begin{itemize}
    \item \textbf{Data}: It consists of the raw and processed data used for model training and inference. The data can include various sources and diverse formats--textual, visual, auditory, or structured. The quality and heterogeneity of data can significantly impact the performance of the LLMs.
    \item \textbf{Model Data and Updates Layer}: This component represents the central hub for the management of the LLMs' operational parameters and their evolution. It tracks adjustments in model architecture and fine-tunes training methodologies, ensuring performance enhancement and seamless adaptability. Beyond these functions, this component is also responsible for managing the model’s resources, including efficient allocation of memory and compute power. It provides a scalable and fault-tolerant interface crucial for serving the LLM's output to downstream applications, which is essential for maintaining the robustness and responsiveness of the system.
    \item \textbf{Prompts}: This component is where prompts are formulated for both inference and learning tasks. These are essentially questions, instructions, or tasks designed to steer the learning of LLMs or elicit specific outputs. Designing well-structured prompts is vital, as it dictates the learning trajectory of the model and the relevance of its responses.   
    \item \textbf{Embeddings}: They interacts with the LLMs inputs and outputs. Specifically, the embeddings translate the prompts and responses to a form the underlying model can understand \cite{chen2018tutorial}. These are high-dimensional vector representations encapsulating the semantic substance of data, enabling the LLMs to process and generate content with contextual significance.
    \item \textbf{Frameworks}: These offer the structure and abstractions necessary for constructing applications powered by LLMs. They orchestrate various components, from LLM providers and embedding models to vector stores and external tools like search engines. Frameworks (e.g., LangChain \cite{langchain}, MetaGPT \cite{hong2023metagpt}, AgentGPT \cite{agentgpt}, AgentVerse \cite{chen2023agentverse}) facilitate the creation of \enquote{chains}, sequences that combine these components, culminating in the critical task of prompting language models \cite{llmsstack}. The crafting of prompts involves integrating information from diverse components with a base prompt template, a pivotal process that dictates the performance and output of LLMs.
\end{itemize}

     \begin{figure*}[ht!]
    \centering
      \includegraphics[width=0.8\textwidth]{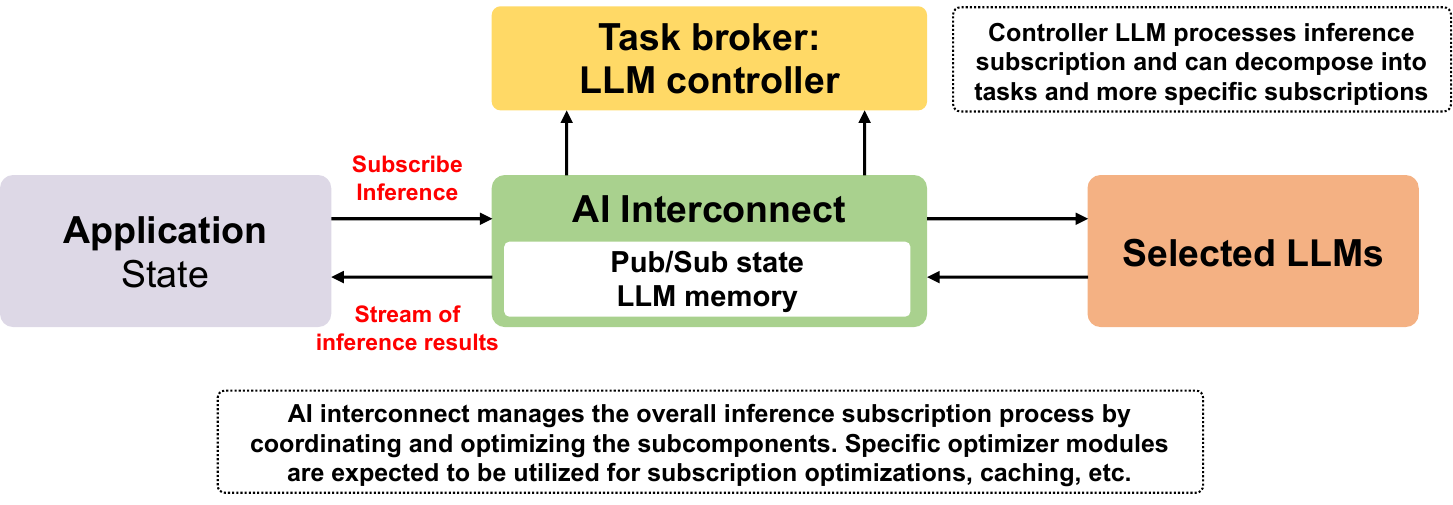}
       \centering
      {\textbf{\caption{Overview of inference subscription, in which AI Interconnect employs a task broker component to analyze and decompose the input into sub-tasks and subscriptions.}\label{fig:7}}}
    \end{figure*}
    
The envisaged API, as depicted in Figure \ref{fig:6}, aims to streamline interactions across these components with several functionalities. Firstly, it enables networking software components to \textit{subscribe} to inference capabilities, seamlessly obtaining AI-generated predictions or results without the need of explicitly requesting them for each instance. Secondly, it allows components to \textit{subscribe} to the learning capabilities, enabling components to align with the latest changes and updates of training data and models. Thirdly, it enables to partake in inference and learning tasks while keeping the input data close to its source (e.g., through the use of network coordinates systems \cite{donnet2010survey}), prioritizing then data locality to reduce latency, conserve bandwidth, and uphold data privacy. Lastly, it allows networking software components to \textit{subscribe} to semantics-aware data communication strategies, prioritizing critical data and ensuring pertinent updates are dispatched promptly and efficiently.

While the envisaged AI Interconnect pub/sub API is simple and extensible through parameter metadata, it may be necessary to request or subscribe to a variety of data and results. The subscription can also be one-shot (\textit{request-notify}). The AI Interconnect can coordinate such processing by requesting or subscribing to a specification detailing the interaction requirements.  

Figure \ref{fig:7} provides an overview of the subscribe inference operation, in which the AI interconnect uses a task broker component to decompose the operation into elementary components that are then executed. The AI Interconnect is responsible for the coordination of the process in the distributed environment. 

The 'participateInference' operation in a pub/sub-API might initially seem counter-intuitive, as pub/sub systems are traditionally designed for asynchronous and decoupled communication between publishers and subscribers of information \cite{19502804fcb7443fa4dfdc54b802471d}. However, the integration of this kind of \textit{request-notify} operation into a pub/sub framework offers several advantages. These include the decoupling of components (the requester does not need to know which server or LLM handles the request), support for load distribution through indirection, flexibility by allowing the exchange and modification of components, and an asynchronous completion token that also supports function callbacks from the LLM system (session-related information is expected to be carried in the parameter metadata). Additionally, it allows for rich semantics in terms of interactions and workflows, and enhanced data flow where both raw data and partial inference results can flow seamlessly across components, facilitating complex pipelines.

The 'participateLearning' operation is akin to 'participateInference' in many ways but is specifically designed for ML and the iterative improvement of models. Its integration within a pub/sub framework can yield multiple benefits:

\begin{itemize}
  \item \underline{Continuous Learning}: In dynamic computing environments, models need to be updated regularly to capture the latest trends or changes in data \cite{kambatla2014trends}. Using 'participateLearning', nodes in a system can continually send new data samples for the model to learn from, facilitating an ongoing training process.
  \item \underline{Decentralized Model Training}: In distributed systems, particularly in IoT or edge computing contexts, devices or nodes might gather data that can be useful for model training \cite{chen2019deep}. These devices can use the 'participateLearning' operation to contribute their data to a centralized or distributed model training process without needing a direct connection to the training server or system.
  \item \underline{Collaborative Learning}: Multiple devices or nodes might have different views or subsets of data. By participating in learning, they can collaboratively contribute to building a more robust and holistic model that captures a broader spectrum of information \cite{lu2019collaborative}. 
  \item \underline{Asynchronous Updates}: Systems can submit data for training without waiting for the training to complete. They are notified via callbacks or other mechanisms when the model has been updated, allowing for non-blocking operations.
  \item \underline{Efficient Resource Utilization}: Not all nodes might be equipped to perform intensive learning tasks in a large-scale distributed systems. By allowing nodes to simply participate in the learning process without directly handling the computational demands of training \cite{luo2021cost}, it possible to ensure that learning happens on suitably equipped nodes (such as cloud servers or dedicated ML training hardware).
  \item \underline{Feedback Loops}: After participating in the learning process, nodes can receive updates about the performance or accuracy of the new model (in a similar fashion as done in \cite{andreina2021baffle}), which can guide further data collection or preprocessing at the edge.
\end{itemize}

In essence, both 'participateInference' and 'participateLearning' operations provide scalable and flexible mechanisms for decentralized and collaborative inference and model training, harnessing the distributed nature of data collection and the asynchronous communication benefits intrinsic to the pub/sub model.

     \begin{figure*}[ht!]
    \centering
      \includegraphics[width=0.7\textwidth]{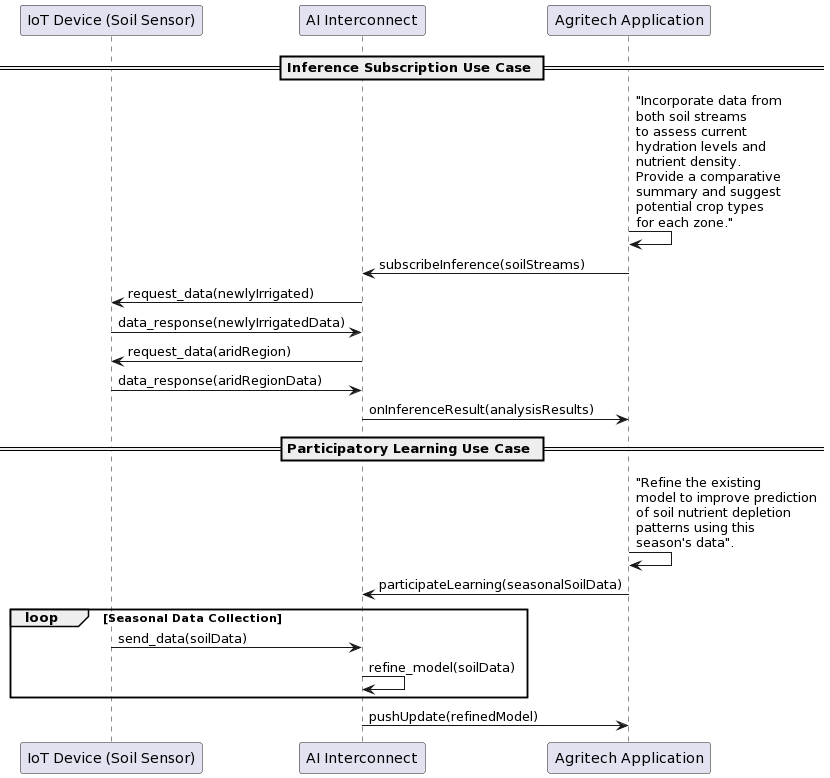}
       \centering
      {\textbf{\caption{UML Sequence Diagram illustrating the interaction flow for Smart Agriculture System use cases. The diagram represents the subscription to real-time inference updates for optimizing agricultural processes (Inference Subscription Use Case) and the collaborative improvement of predictive models through participatory learning from diverse data points (Participatory Learning Use Case).}\label{fig:agriculture}}}
    \end{figure*}
    
The AI Interconnect plays a crucial role in selecting the appropriate pub/sub or message bus for specific interactions, employing an LLM to analyze interaction requirements and choose the most suitable configuration. This selection considers various factors, such as the participating components, real-time responsiveness, and data nature. The ability to select the most suitable pub/sub and request/reply API is critical for enhancing system interoperability, allowing it to integrate seamlessly with various O-RAN implementations such as the RIC Message Router (RMR) \cite{O-RAN-RMR}, as well as with broader distributed event streaming platforms like Apache Kafka \cite{Kafka}.

Ultimately, the selection of the messaging fabric ensures the use of proper and interoperable technological solutions. In conjunction, sandboxing and scoping techniques ensure that message flows are application-specific while permitting the sharing of messages across applications when necessary \cite{tarkoma2009publish}.

To demonstrate the versatility of the AI Interconnect, we present a general-purpose application scenario involving a Smart Agriculture System, where a network of IoT devices is deployed across a large farm to monitor and optimize agricultural processes, distinct from the network-centric examples explored in the following sections. The main controller interfaces with sensors and actuators using the pub/sub API mechanism. Through this mechanism, the controller can subscribe to or query data streams and initiate specific tasks based on the insights generated by the LLMs.

\textbf{Inference Subscription Use Case:} An agritech application requires real-time insights from two different soil data streams—one from a newly irrigated zone and another from an arid region. Rather than manually processing this information, the application issues a subscription request, through the \textit{subscribeInference} method, to the AI Interconnect with the following prompt:

\textcolor{blue}{\enquote{Incorporate data from both soil streams to assess current hydration levels and nutrient density. Provide a comparative summary and suggest potential crop types for each zone.}}

Subsequently, the AI Interconnect continuously provides updates, through the \textit{onInferenceResult} method, based on its ongoing analysis to the subscriber as the data from the streams changes.

\textbf{Participatory Learning Use Case:} Consider an application deployed across various farms that has gathered a wide array of data points. Looking to enhance its existing model for greater accuracy in future predictions, the application utilizes the AI Interconnect by pushing the accumulated data over the season:

\textcolor{blue}{participateLearning (seasonal-soil-data, \enquote{Refine the existing model to improve prediction of soil nutrient depletion patterns using this season's data.})}

In response, the AI Interconnect employs this data to fine-tune and bolster the model, which the application can then utilize for more accurate predictions in the future.

     \begin{figure*}[ht!]
    \centering
      \includegraphics[width=0.7\textwidth]{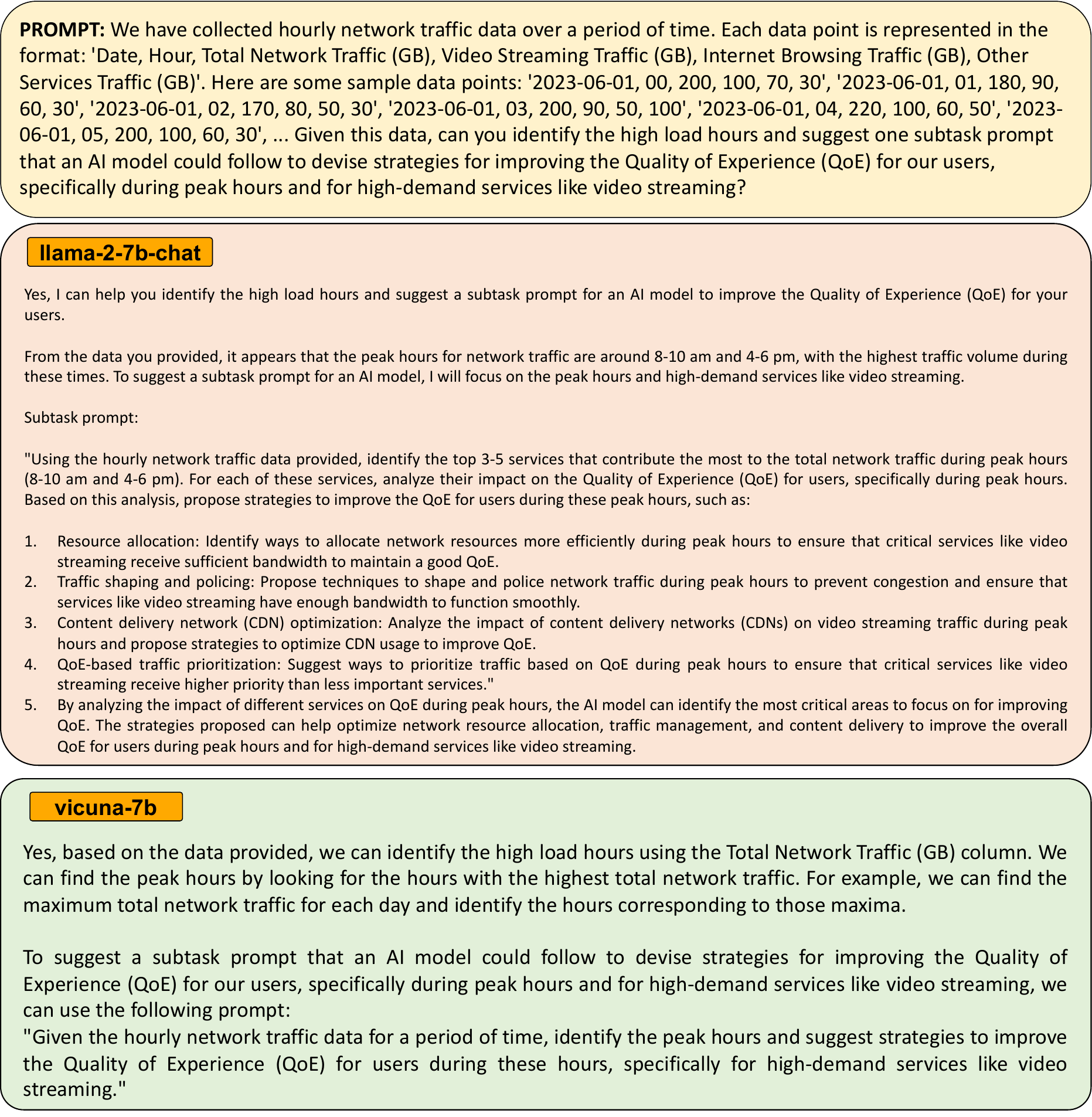}
       \centering
      {\textbf{\caption{This example illustrates LLMs' potential for serving as network controllers. It presents a basic prompt with network usage data, requesting network optimization through sub-prompts. Responses from two well-known LLM models, LLaMA2-7B and Vicuna-7B, are shown. This detailed one-to-one comparison is performed through \cite{zheng2023judging}.}\label{fig:8}}}
    \end{figure*}

The AI Interconnect demonstrates its capability to use LLMs not only for parsing high-level language prompts into actionable insights but also for orchestrating the right distributed event streaming methods, ensuring seamless service provisioning as illustrated in Figure \ref{fig:agriculture}.

\section{LLMs as Controller}\label{sec:8}
The integration of LLMs as controllers in network and application domains is gaining momentum, leveraging their capacity for advanced tasks management \cite{mani2023enhancing}. They can effectively decompose user intent into manageable tasks, enabling a more efficient and targeted approach to problem-solving \cite{netsoftintent}. LLMs can select appropriate functions based on function templates, which provide a structured framework for task execution. This selection process is guided by the LLM's understanding of the task requirements and the capabilities of the available functions. Once the tasks have been defined and the functions selected, the LLM guides the execution of these tasks to completion. This process involves monitoring task progress, managing resources, and making adjustments as necessary to ensure successful task completion. Using LLMs as controllers thus brings a high level of intelligence and adaptability to network and application task management.

Metaprompting emerges as a technique where an initial prompt generates a series of subsequent prompts, which are then processed by other agents or components \cite{reynolds2021prompt}. This feature can be particularly useful in complex systems where a single initial prompt might not be sufficient to describe a problem or task fully. By generating a series of metaprompts, the system can explore different aspects of the problem or task in a structured and systematic way.

     \begin{figure*}[ht!]
    \centering
      \includegraphics[width=0.8\textwidth]{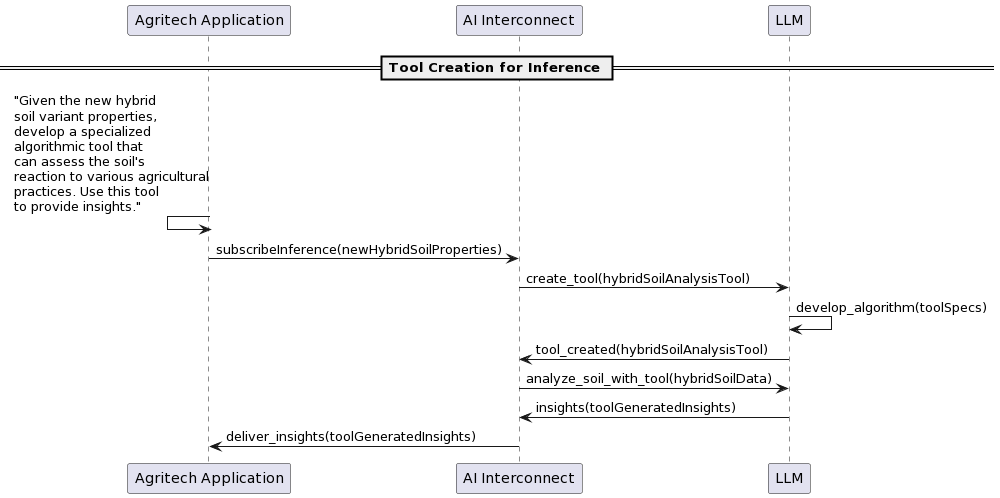}
       \centering
      {\textbf{\caption{The AI Interconnect communicates with the underlying LLM to create a specialized tool for analyzing the hybrid soil variant. The LLM develops the algorithm (the tool) and provides the insights back to the AI Interconnect, which in turn delivers these insights to the Agritech Application.}\label{fig:toolcreation}}}
    \end{figure*}
    
Within our envisioned architecture, control flows associated with LLMs can be orchestrated through two principal approaches. The initial strategy involves management at the network element level, harnessing the standard operational procedures embedded within the network hardware to oversee LLM-related interactions. This ensures that the implementation of LLMs as controllers is seamlessly integrated with the existing network infrastructure, maintaining system integrity while injecting advanced AI-driven capabilities. 
   
Alternatively, the LLM-related control flow can be partially offloaded to the AI Interconnect alongside other LLMs, harnessing the asynchronous capabilities of publish/subscribe mechanisms and callbacks. This arrangement positions the AI Interconnect as a dynamic mediator, orchestrating the exchange between LLMs and network entities. Asynchronous callbacks, in particular, enable operations to proceed uninterrupted, fostering higher levels of system concurrency and significantly boosting efficiency and responsiveness.

Employing a distributed placement mechanism for control logic is highly beneficial in managing the intricate and demanding operations of LLMs. By disseminating control responsibilities, the system mitigates network strain and capitalizes on the AI Interconnect and other LLMs' computational capabilities. As highlighted in \cite{liu2023bolaa}, this distributed approach augments scalability and agility, thereby enhancing system performance and flexibility.

LLM orchestration and optimization can also be realized through the distribution of models, employing strategies like split learning \cite{gupta2018distributed}. This enables the execution of portions of the model closer to the data source, potentially on-device, aligning with the edge-cloud continuum. While local execution is preferred for privacy and security reasons, it may present trade-offs in terms of performance and scalability, which necessitates a balanced approach to model distribution.

To illustrate LLM's potential for serving as network controllers, Figure \ref{fig:8} presents a basic prompt with network usage data requesting network optimization through sub-prompts. The prompt response is presented for two well-known LLM models, Vicuna-7B \cite{chiang2023vicuna} and LLaMA2-7B \cite{touvron2023llama}. The sub-prompts indicate directions for further analysis; however, they are both too high-level and distinctly varied to support network optimization process automation. Further elaboration would be needed for extracting fine-grained sub-prompt responses from the LLM. The example illustrates the potential benefit of LLMs, also models running on edge nodes, for network management.

\section{LLMs as Dynamic Tool Builder and API Broker}\label{sec:9}

LLMs possess have the innovative capacity to serve as dynamic tool selectors and generators, introducing increased adaptability and intelligence within system operations. In their role as tool selectors, LLMs scrutinize task demands to pinpoint the most fitting tools or APIs for the job. This process leverages their deep comprehension of available tools' capabilities, ensuring alignment with the task objectives \cite{ruan2023tptu, qian2023creator}. Conversely, LLMs demonstrate versatility as tool generators by creating or tailoring tools to suit particular tasks, which could entail formulating new APIs or refining the configurations of existing ones for enhanced efficacy \cite{patil2023gorilla}. This versatile nature of LLMs as selectors and creators amplifies system flexibility and efficiency, empowering systems to adapt seamlessly across a diverse array of tasks and operational challenges.

\subsection*{Example 1: Smart Agriculture Tool Creation}

Following our agriculture use case examples; our application is confronted with a novel case--the introduction of a new hybrid soil variant in a controlled setting. Given that conventional soil analysis tools may not be up to the task, the application invokes assistance from the AI Interconnect with the following prompt:

\textcolor{blue}{\enquote{Given the new hybrid soil variant properties, develop a specialized algorithmic tool that can assess the soil's reaction to various agricultural practices. Use this tool to provide insights.}}

In response (Figure \ref{fig:toolcreation}), the AI Interconnect, utilizing underlying LLM capabilities, not only constructs the requested tool but also ensures that the application receives the tool-generated insights. It is important to note that while the development of on-the-fly tool creation is in its early stages, the process necessitates rigorous human oversight and comprehensive auditing to verify the accuracy and reliability of the tools and insights produced.

\subsection*{Example 2: Dynamic Network Orchestration for O-RAN}

Part of O-RAN's design includes the adoption of pub/sub mechanisms to facilitate asynchronous and real-time communication between different functional entities in the RAN \cite{polese2023understanding}.
Within the O-RAN architecture, xApps are applications that run on the non-RealTime (Non-RT) RIC. These applications add value by creating policies that the Non-RT RIC delivers to the near-RealTime (Near-RT) RIC through the A1 interface, handling non-real-time tasks like long-term network optimization and planning \cite{oran_nonrt}. 

\begin{figure*}[ht!]
     \centering
     \begin{subfigure}{\textwidth}
         \centering
         \includegraphics[width=0.7\textwidth]{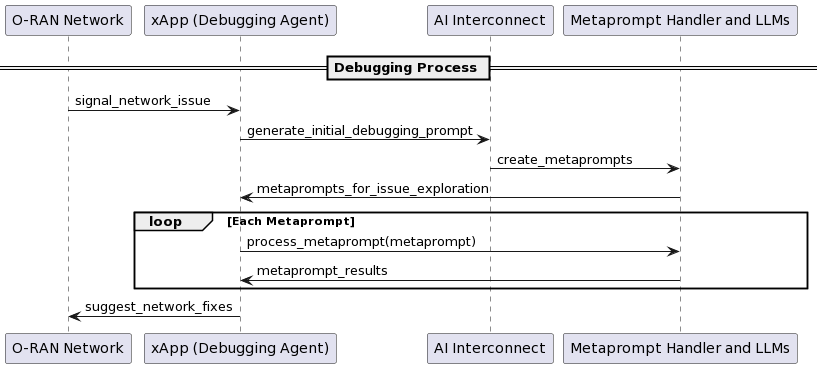}
         \caption{}
         \label{fig:orandeb}
     \end{subfigure}
     \hfill
     \begin{subfigure}{\textwidth}
         \centering
         \includegraphics[width=0.7\textwidth]{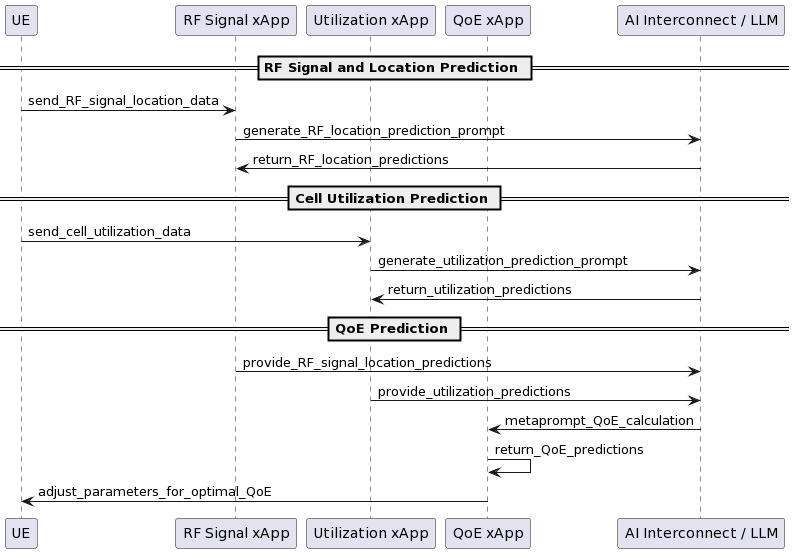}
         \caption{}
         \label{fig:oranmulti}
     \end{subfigure}
     \caption{Sequence diagram illustrating the application of LLMs and metaprompting within xApps to debug network issues (a) and predict network and user equipment behavior to enhance QoE (b) in the O-RAN architecture.}
\end{figure*}

In the context of O-RAN and xApps, AI models can serve as agents to consume data, generate prompts, and emit results. These agents, placed within xApps, perform tasks such as analyzing network data, identifying issues, and generating task prompts to guide operations.  In this setup, metaprompting could guide the debugging process. For example, an initial debugging prompt might identify a potential issue with the network. This prompt could then generate a series of metaprompts, each exploring a different aspect of the issue. These metaprompts could then be consumed by different agents or components within the xApp, allowing them to investigate the issue from different angles and potentially identify a solution (Figure \ref{fig:orandeb}).

GPT models and metaprompting can provide a powerful and flexible way to manage and optimize network operations within an O-RAN context. For instance, an xApp application could consume measurements of radio frequency (RF) signal experienced by a UE and its location and output future predictions of the UE’s location and the RF signal at that location. Another xApp might use cell utilization measurements and output future predictions of cell site utilization. Yet another xApp might consume outputs from other xApps and actual measurements to calculate the potential Quality of Experience (QoE) for a UE (Figure \ref{fig:oranmulti}).

The 'AbstractPubSubAPI' is envisioned as a unified interface that harmonizes with O-RAN's pub/sub mechanisms. This API enhances service model and information element management, allowing RAN entities to exchange data efficiently. It empowers the RAN RIC to dynamically refine algorithms through continuous learning and provides comprehensive methods for both near-real-time and non-real-time RAN control. Furthermore, the API underscores orchestrated management and automation, integrating with Service Management \& Orchestration (SMO) components for efficient operation. It also aims to support fault management, self-healing, and RAN slicing, facilitating real-time performance adjustments based on AI inferences.

\section{Training Foundation LLMs for Mobile Networks Applications: Introducing TelecomLLM}\label{sec:7}

Building on the conceptual principles seen so far, we envision the creation of a \textbf{TelecomLLM}. This LLM would be specifically tailored to the telecommunications domain, integrating domain-specific knowledge, algorithms, and methodologies. Such a model would not only possess an inherent understanding of the intricacies of the telecom networks, protocols, and standards but also offer advanced capabilities for applications such as traffic forecasting \cite{feng2022adaptive, ferreira2023forecasting}, anomaly detection \cite{siriwardhana2021ai}, network automation \cite{coronado2022zero}, and code generation \cite{mani2023enhancing}. For instance, models dedicated at understanding and generating code can streamline and automate the development and optimization of network protocols and algorithms.
The potential of these models in enhancing the intelligence and efficiency of wireless networks is being recognized and embraced. Researchers and practitioners have already started exploring the use of LLMs and GPTs in these domains, further reinforcing their prospective value.

Clearly, developing LLMs and GPTs specifically tailored for mobile networks applications presents a unique set of challenges and considerations. Some of these are inherent to the telecom domain itself, while others are associated with the intricacies of model development. In the context of the telecom domain, understanding the specific needs and constraints of mobile networks becomes imperative. These systems are inherently dynamic, potentially resource-constrained, and highly susceptible to environmental factors. On the model development front, \textit{Data Volume and Diversity}, \textit{Model Architecture}, and \textit{Model Update} play a pivotal role. 

\subsection*{Data Volume and Diversity:} Training a TelecomLLM necessitates a substantial volume of data to capture the diverse and dynamic nature of telecommunication networks. The exact amount of data needed depends on the complexity and depth of tasks the model is expected to perform. Generally, more data offers better performance, but with diminishing returns after a certain point. Mobile networks generate vast amounts of data, varying not only in volume but also in type. Ensuring that the model training data is comprehensive and reflects this diversity is crucial. Therefore, the training data should encompass a broad spectrum of network scenarios, use cases, traffic patterns, user behaviors, and potential anomalies. It is also crucial to include data under various environmental conditions, different equipment types, and from different geographical regions to build a model that generalizes well.

\subsection*{Model Architecture:}
The choice between a single massive GPT-type model and multiple smaller, specialized models depends on the deployment scenario and specific use cases.

\begin{itemize}
    \item \textbf{Single Massive Model:} A large, comprehensive GPT-type model may offer broad capabilities, understanding various telecommunication protocols, standards, and technologies. It may be suitable for a wide range of tasks, offering a ‘one-size-fits-all’ solution. However, it might also require substantial computational resources for training and inference and might not be optimal for specific, specialized tasks.
    \item \textbf{Multiple Specialized Models:} Having multiple smaller models specialized for specific tasks or domains within telecommunications may offer more efficiency and accuracy for those tasks. These models would be quicker to train, more computationally efficient to run, and could be fine-tuned more easily for particular applications. This approach would require a framework for managing and deploying multiple models efficiently.
\end{itemize}

The decision may be a balance between the two approaches, possibly deploying a large general model for common, broad tasks, and specialized models for niche or critical applications. The trade-off between generalization and specialization, along with considerations of computational efficiency and deployment flexibility, should guide the architecture design for TelecomLLM. Regardless of this choice, we foresee the emergence of a TelecomLLM marketplace, which can facilitate the development for both specialized and general massive models. This marketplace can serve as a hub for developers, researchers, and telecom experts to access, contribute to, or procure specialized TelecomLLMs suited to specific tasks or domains. Central to the success of such a marketplace are standardization efforts and coordinated open-source initiatives. These efforts not only ensure interoperability and compatibility but also foster a collaborative environment for the continual advancement and refinement of TelecomLLMs.

\begin{figure*}[!htbp]
     \centering
     \begin{subfigure}{\textwidth}
         \centering
         \includegraphics[width=0.7\textwidth]{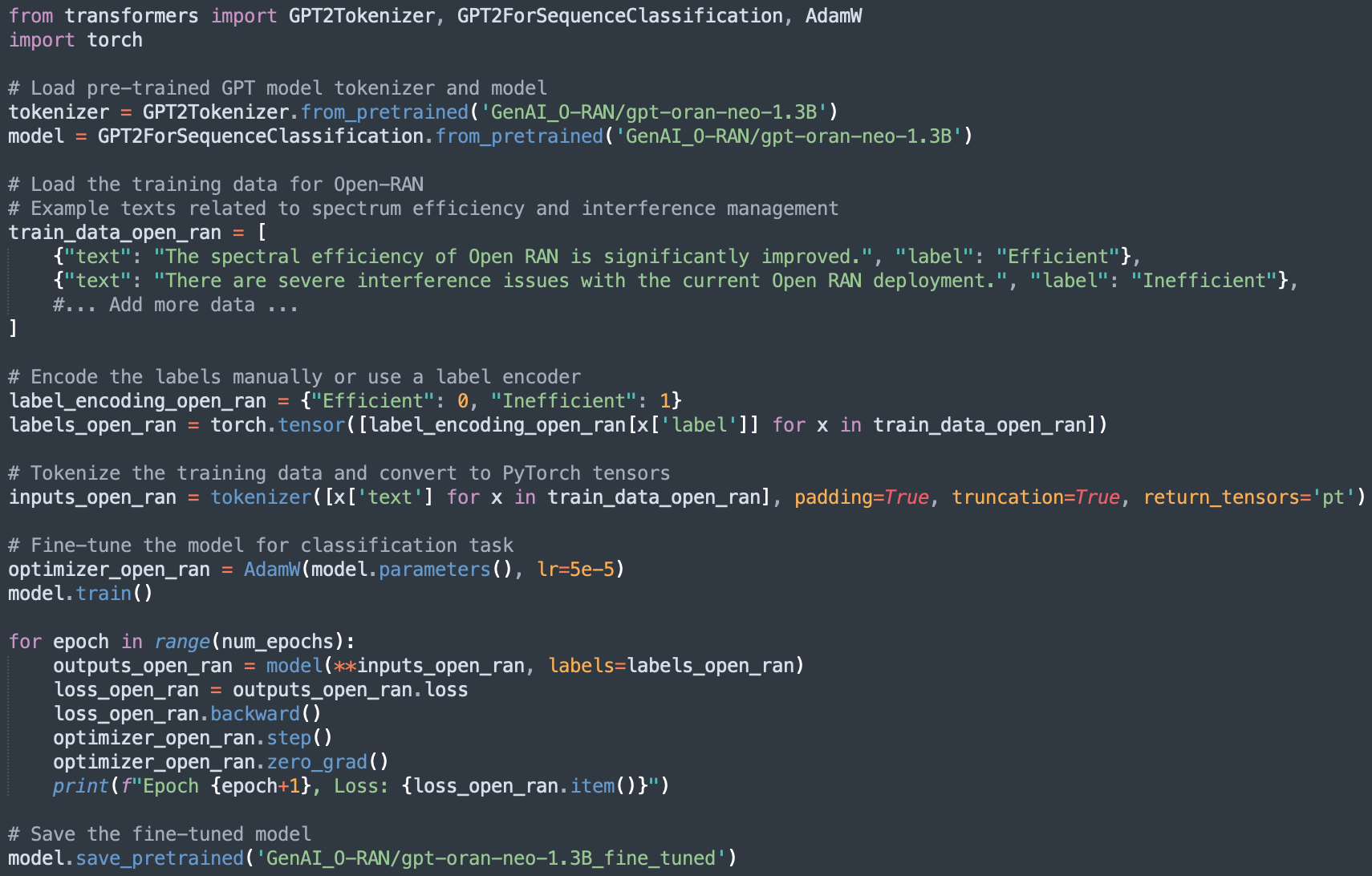}
         \caption{}
         \label{fig:oran_finetune}
     \end{subfigure}
     \hfill
     \begin{subfigure}{\textwidth}
         \centering
         \includegraphics[width=0.7\textwidth]{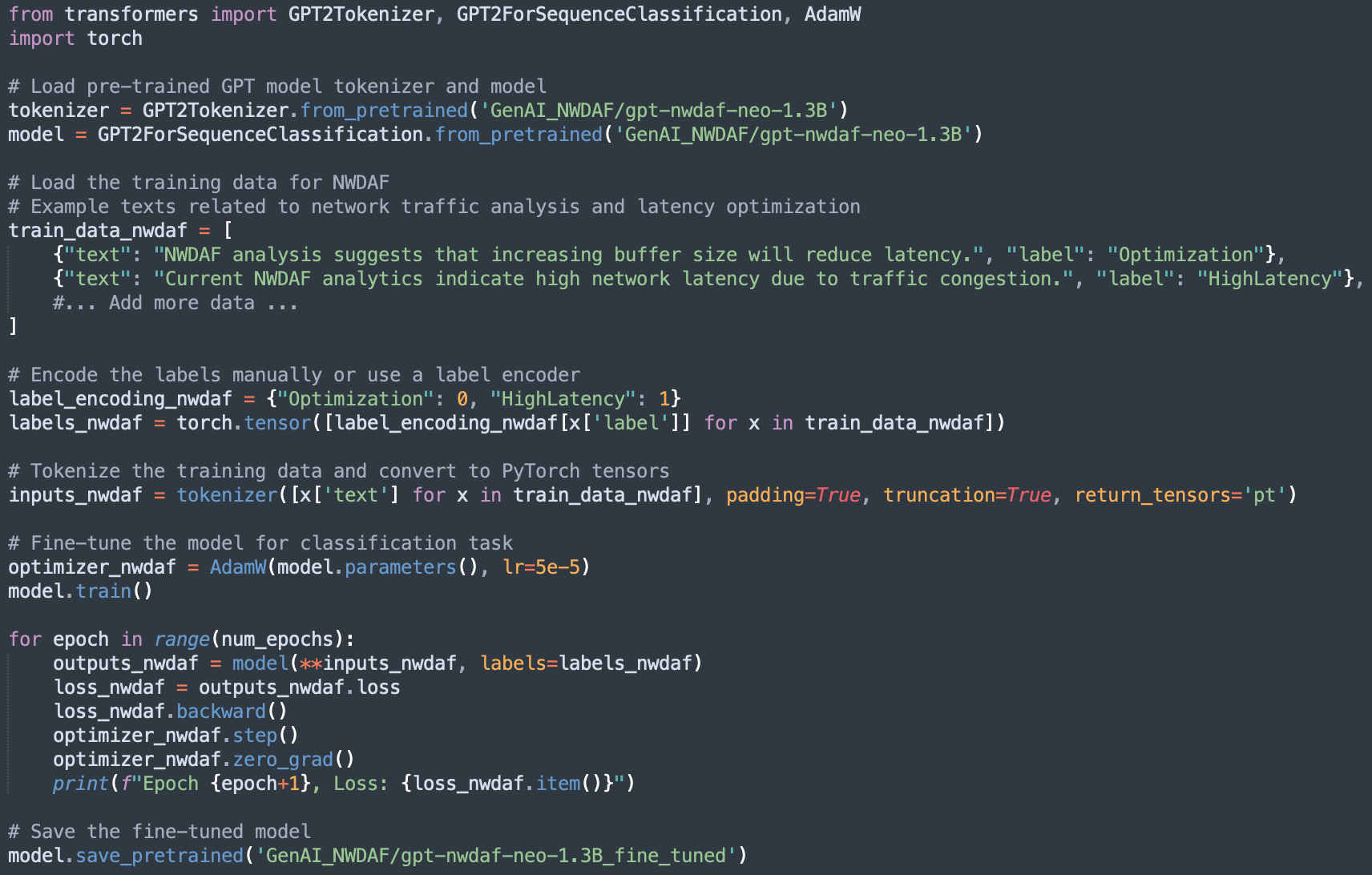}
         \caption{}
         \label{fig:nwdaf_finetune}
     \end{subfigure}
     \caption{Exemplified code snippets for the fine-tuning of a pre-trained GPT model for: Spectral efficiency and interference management using O-RAN-related textual data (a); Network traffic analysis and latency optimization using NWDAF-related textual data (b).}\label{fig:oran_finetune.}
\end{figure*}

\subsection*{Model Update:} Considering the rapidly evolving landscape of telecommunications, an adaptive framework for \textit{continuous learning} and updating of the models is crucial. This approach allows the TelecomLLM to stay current with the ongoing changes and innovations in the field, improving its reliability and usefulness over time. In this respect, several existing LLMs and GPT models can be adapted to support wireless network applications with appropriate training and fine-tuning. 

At its core, fine-tuning involves adjusting one or more internal parameters of a pre-trained model, typically the weights. In the context of LLMs, this process effectively transitions a broad-based model, such as GPT-3, into a specialized model optimized for a distinct application, like ChatGPT \cite{brown2020language}. A significant advantage of this technique is the ability of the models to exhibit enhanced performance. This higher performance comes without the need for vast volumes of manually labeled examples--a common requisite for models predominantly reliant on supervised training.
Opting to fine-tune serves a dual purpose. Firstly, it amplifies the capabilities of the base model. Secondly, a fine-tuned model, even if it is smaller in scale, can often surpass the capabilities of much larger, computationally-intensive models concerning tasks it has been tailored for \cite{ouyang2022training}. 

Fine-tuning an LLM, much like other ML model, begins with the collection of pertinent data, followed by preprocessing, and culminates with model evaluation and deployment. While these steps resonate with conventional ML practices, the unique intricacies of fine-tuning become relevant primarily during the model adjustment phase. Within the context of the LLM, the key of fine-tuning lies in the adaptation of a pre-existing model to better serve a specific use case. This is typically achieved through transfer learning, where a pre-trained model like GPT acts as the foundational structure \cite{zhao2023survey}. By retaining most of the pre-trained weights and only adjusting a subset, transfer learning capitalizes on the existing knowledge of the model while tailoring it to new tasks. This method offers a middle ground, balancing the computational savings against the need to ensure model adaptability. However, there are additional aspects that must be considered. Entirely retraining the parameters, while computationally very demanding, can offer finer-grained control over the model. Transfer learning, despite its advantages, cannot help solving all challenges. For instance, it might struggle with \textit{catastrophic forgetting}, a glitch where neural networks inadvertently erase or overlook previous learning while assimilating new information \cite{zhao2023survey}. 

Although emerging strategies aim to address these possible issue, delving deeper into such techniques remains outside our current discussion's scope. Nevertheless, a rigorous investigation of these methodologies is pivotal if we aspire to develop TelecomLLMs that can be confidently and safely employed in production environments.

The examples shown in Figure \ref{fig:oran_finetune} and Figure \ref{fig:nwdaf_finetune} provide a tangible glimpse into the practical applications of this fine-tuning process, particularly within the TelecomLLM context. The Python code snippets use the Transformers library from Hugging Face \cite{transformers_hugging} to fine-tune GPT-3 on a text classification task related to O-RAN and NWDAF respectively.

In these exemplified code snippets, we start by loading the pre-trained GPT-3 tokenizer and model via the Transformers library. Following this, we delineate a text classification task, sourcing training data pertinent to: \emph{(i)} spectral efficiency and interference for O-RAN (Figure \ref{fig:oran_finetune}), and \emph{(ii)} network traffic analysis coupled with latency optimization for NWDAF (Figure \ref{fig:nwdaf_finetune}). Subsequent steps involve tokenizing the training data with the tokenizer and translating it into PyTorch tensors. The GPT-3 model is then fine-tuned on the classification task, leveraging the AdamW optimizer \cite{adamw} over three epochs. Finally, we save the fine-tuned model to disk. We can then use this model to make predictions by loading it as needed.

In the rapidly advancing field of LLM training, alongside traditional fine-tuning, new methodologies like Retrieval-Augmented Generation (RAG) or Retrieval-Enhanced Transformer (RETRO) are gaining prominence \cite{lewis2020retrieval, borgeaud2022improving}. RAG and RETRO can enhance the LLMs' text generation by including a retrieval system that sources relevant information from a external sources, essentially allowing the model to include external data to enrich its responses. This can improve the model's accuracy and reliability by providing contextually rich background data for generating answers \cite{ragvsfinetune2}.

While we will not delve into the details of these techniques, it is essential to acknowledge the emerging studies that consider their relative merits. The decision to use RAG, RETRO, fine-tuning, or a combination of these is contingent upon the unique demands and goals of the TelecomLLM's application. The choice is dictated by various factors such as the necessity for external data, the desired adaptability of the model's behavior, the dynamics of the data involved, task requirements, and the transparency of the results. In some scenarios, a hybrid approach that integrates both RAG and fine-tuning could yield the best outcomes, leveraging the strengths of both techniques to achieve the desired level of performance and flexibility in the LLMs \cite{ragvsfinetune}.

In conclusion, training a TelecomLLM requires careful consideration of data requirements, strategic decisions on model architecture, and suitable methodology for the model updates. The selection between a single massive model or multiple specialized ones should align with the operational needs and specific use cases in the telecommunications sector, finding a balance between breadth and depth of capabilities, efficiency, and adaptability to the ever-changing telecom environment. 

\section{LLMs Configuration and Interoperability in 6G Networks}\label{sec:10}

In distributed and networked environments that utilize GPT/LLM models, the interaction and cooperation between various entities require a unified basis for communication \cite{chen2023netgpt, liu2023bolaa, hong2023metagpt}. This is where aspects like the configuration of AI  models and interoperability become crucial. Furthermore, considering the complex context of 6G networks, the importance of these factors transcends simple data translation. Model configuration and interoperability emerge as fundamental elements that not only facilitate various aspects of network functionality but also contribute to the governance and adaptability of the system \cite{hossain2022ai}.

\begin{figure*}
     \centering
     \begin{subfigure}{\textwidth}
         \centering
         \includegraphics[width=0.8\textwidth]{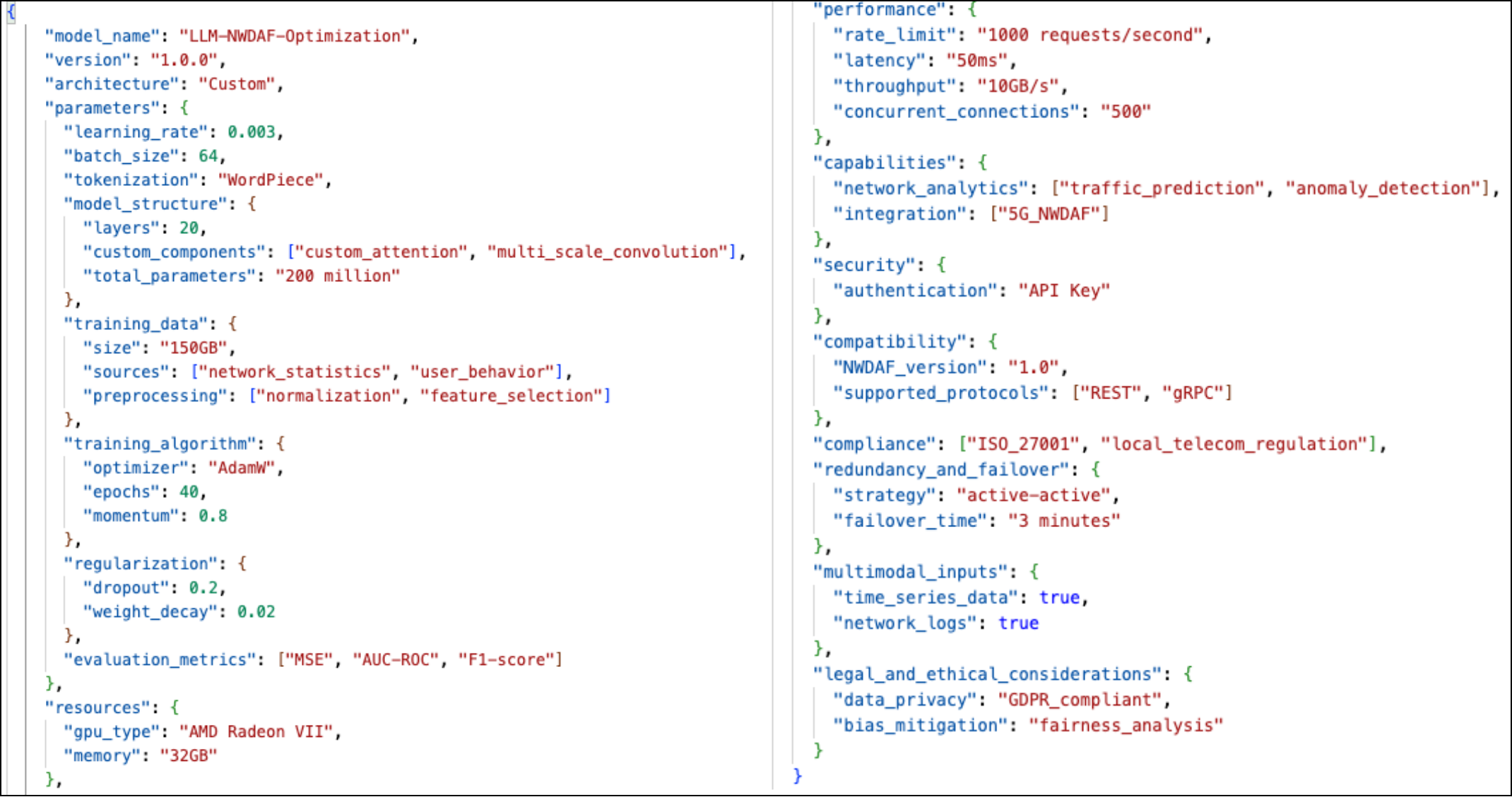}
         \caption{}
         \label{fig:14}
     \end{subfigure}
     \hfill
     \begin{subfigure}{\textwidth}
         \centering
         \includegraphics[width=0.8\textwidth]{figures/nwdaf_config_new.pdf}
         \caption{}
         \label{fig:15}
     \end{subfigure}
     \caption{Example of GPT Configuration file in the NWDAF (a) and O-RAN (b) cases. The file details the GPT/LLM model essentials, from architecture and hyper-parameters to 6G-specific capabilities and security measures.}
\end{figure*}

Data serialization is a key enabler for ensuring seamless and consistent model configuration and interoperability \cite{sumaray2012comparison}. It involves translating complex data structures, including model parameters, into a format that can be easily stored, transmitted, and reconstructed. This process is crucial in a 6G environment, filled with myriad devices, platforms, applications, and network components, particularly when different nodes or components employ disparate GPT/LLM models with varying specifications. Serialization ensures that data can be consistently interpreted, irrespective of underlying variations, paving the way for different GPT/LLM models to communicate and work together seamlessly. In the context of 6G, data serialization is not merely a technical procedure, but it is an essential enabler that helps networks achieve the flexibility, responsiveness, and coherence required to fulfill the promises of LLMs-enabled next-generation mobile networks. Whether it is enhancing interoperability, efficiency, ease of configuration, or dynamic adaptation, serialization acts as a vital link connecting different networked elements, helping them function as a cohesive and intelligent whole.

In the following, we will dig deeper into the requirements and varied scenarios concerning LLMs configuration and interoperability within the context of 6G. From the definition and alignment of diverse model capabilities to the seamless exchange of GPT-linked information across heterogeneous network components, we will try to shed some light on practical operational scenarios that will shape the intersection of GPT technologies (as a reference example for LLMs) and 6G networks.

\subsection*{6G-LLMs Configuration.} Standardized model configuration is a key prerequisite for utilizing GPTs/LLMs in various 6G components like NWDAF, RIC, etc. This standardization allows for uniformity, ease of integration, and adaptability to specific use cases. A crucial aspect of this standardization is the use of a data serialization language that is both human-readable and machine-readable.
Two popular options for data serialization are JavaScript Object Notation (JSON) \cite{json}, known for being lightweight and widely supported, and YAML Ain't Markup Language (YAML) \cite{yaml}, recognized for supporting complex data structures \cite{eriksson2011comparison}. While JSON is quick to parse but lacks support for certain data types configurations \cite{eriksson2011comparison, baazizi2019schemas}, YAML, though more \enquote{human-friendly}, can be more prone to errors if not formatted correctly. 

     \begin{figure*}[htb]
    \centering
      \includegraphics[width=0.8\textwidth]{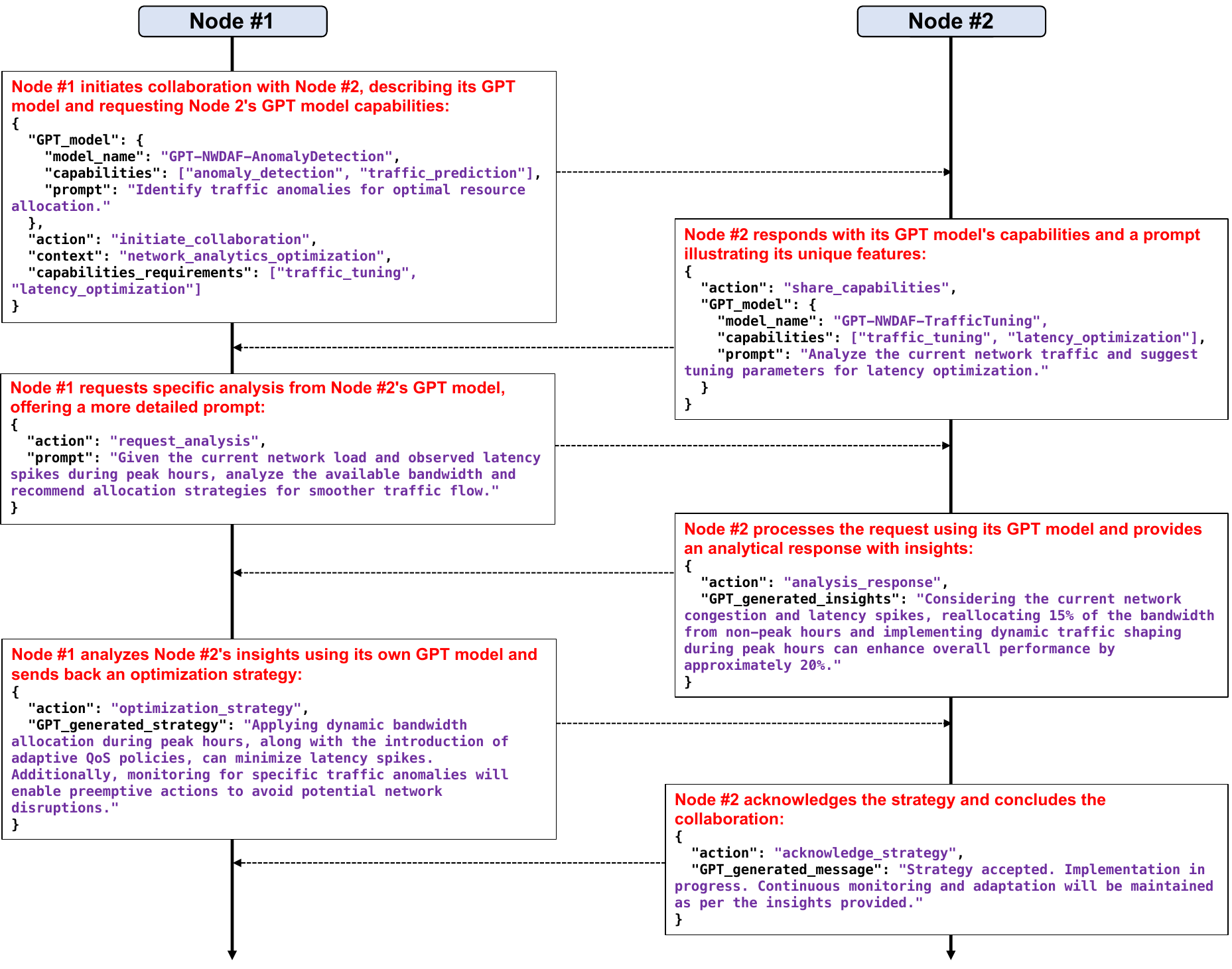}
       \centering
      {\textbf{\caption{GPT-powered Nodes interaction in NWDAF. The example showcases two distinct nodes interacting seamlessly for enhanced network optimization, demonstrating the potential of tailored AI capabilities in varied scenarios.}\label{fig:9}}}
    \end{figure*}
    
In the context of our examples, we choose to proceed with JSON. However, it is worth noting that YAML and JSON can both serve distinct purposes effectively. YAML is frequently chosen for setting up infrastructure and implementing \enquote{infrastructure as code} strategies due to its user-friendly readability. In contrast, JSON is commonly preferred for web APIs owing to its robust support for data serialization \cite{jsonvsyaml}.

When describing GPT/LLM models across different computing units, it is essential to include information regarding the model's type, version, and underlying architecture details \cite{zhao2023survey}. Specifics about the underlying algorithm, such as Transformer architecture, are key, as are hyperparameters like learning rate and batch size. Preprocessing settings, training data details, hardware specifications, performance metrics, and availability are also critical components. In addition, the model's capabilities and features need to be detailed, including supported domains specific to 6G networks or applications functionality, along with any extensions or plugins available. Security and compliance considerations, such as authentication methods, encryption standards, and privacy policies, must be meticulously addressed.
In the following, we envision how a GPT configuration file for NWDAF and O-RAN functionality could look like, integrating all these aspects into a coherent and standardized form. The goal is to provide a concrete framework that informs the development, integration, and utilization of GPTs/LLMs in the emerging 6G landscape, ensuring the thorough outline of all key aspects and requirements.
For instance, an LLM tailored for NWDAF might be configured as shown in Figure \ref{fig:14}. Similarly, a GPT designed to facilitate resource allocation within O-RAN might look like in Figure \ref{fig:15}.
    
The configuration of LLMs for specific 6G functionality, such as NWDAF optimization and O-RAN resource allocation, signifies a detailed integration of architecture, learning parameters, resources, security, performance, and more. In the NWDAF case, emphasis on network analytics and custom architectural elements supports tasks like traffic prediction and anomaly detection. For O-RAN, the focus shifts to efficient resource allocation, radio network optimization, and integration with O-RAN standards. In both scenarios, such configurations encapsulate essential details, from the models’ structure and capabilities to security measures and performance metrics, including rate limits, latency, throughput, and concurrent connections. This explicit and standardized definition streamlines deployment, scale, and interoperability, paving the way for consistent and efficient application of AI technologies across the 6G landscape.

\subsection*{6G-LLMs Interoperability.} Beyond mere configuration, the dynamic nature of 6G networks demands robust interoperability among heterogeneous nodes. Central to this requirement is the AI Interconnect's role in overseeing operations, managing data flows, and controlling network traffic. It ensures that different nodes, each running their own LLM/GPT models, can establish a common language or protocol to collaborate seamlessly on tasks such as network optimization or fault prediction.

     \begin{figure*}[htb]
    \centering
      \includegraphics[width=0.8\textwidth]{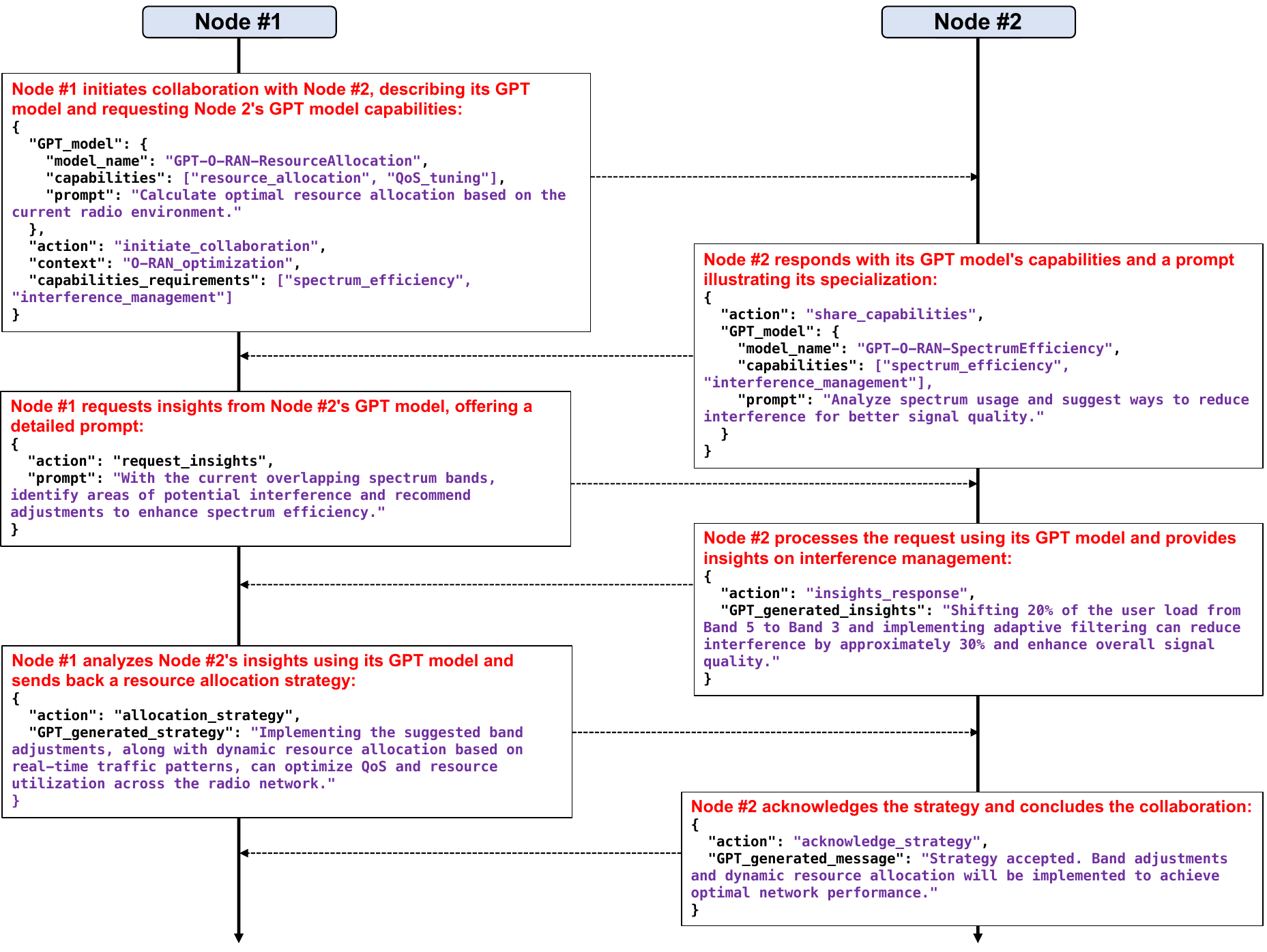}
       \centering
      {\textbf{\caption{GPT-powered Nodes interaction in O-RAN. The example showcases two distinct nodes interacting seamlessly for negotiating about spectrum efficiency, resource allocation, and QoS tuning.}\label{fig:10}}}
    \end{figure*}
    
Facilitated by the AI Interconnect, nodes can engage in a coherent interactions through a sequence of prompts, intents, and responses, underpinned by a shared data format and APIs. For instance, one node might issue a predictive request for resource allocation in anticipation of traffic loads, while another node, upon receiving this request, would allocate resources accordingly, employing the mutually understood algorithm. This dynamic interplay, expertly orchestrated by the AI Interconnect, allows nodes to adapt to abrupt network variations, renegotiate metrics, and maintain uninterrupted collaboration, thereby underscoring the imperative of interoperability in a 6G ecosystem.

In the following, our focus will shift towards visualizing how future 6G Network entities can set up and fine-tune GPT-enabled interoperable tasks, still in the context of NWDAF and O-RAN functionality. The possibility of seamless interaction and collaboration will be explored across different scenarios and cases.

The example shown in Figure \ref{fig:9} involves two NWDAF entities, each using different GPT models. The interaction shows network optimization capabilities, using the specific capabilities of each GPT model.

This exchange demonstrates the capabilities of using different GPT models within the NWDAF context. The verbose prompts and responses highlight the ability of these models to instruct the underlying nodes' components of complex analytical tasks and communicate insights and strategies effectively. By utilizing distinct GPT models that are specialized in different aspects of network optimization, the nodes are able to collaborate on a more sophisticated level enhancing the network’s adaptability and efficiency.

Similarly, the example of Figure \ref{fig:10} shows an interaction between two nodes within the context of O-RAN. These nodes are using two GPT models to collaboratively work on radio network optimization and QoS tuning.

This scenario emphasizes the collaborative capabilities of different GPT models within the O-RAN context. The prompts and responses illustrate how these models can engage in complex interactions regarding spectrum efficiency, resource allocation, and QoS tuning. By leveraging the capabilities of each GPT model and their respective underlying systems, the nodes are able to generate robust strategies for optimizing the RAN.

\begin{figure*}[htb]
    \centering
      \includegraphics[width=0.8\textwidth]{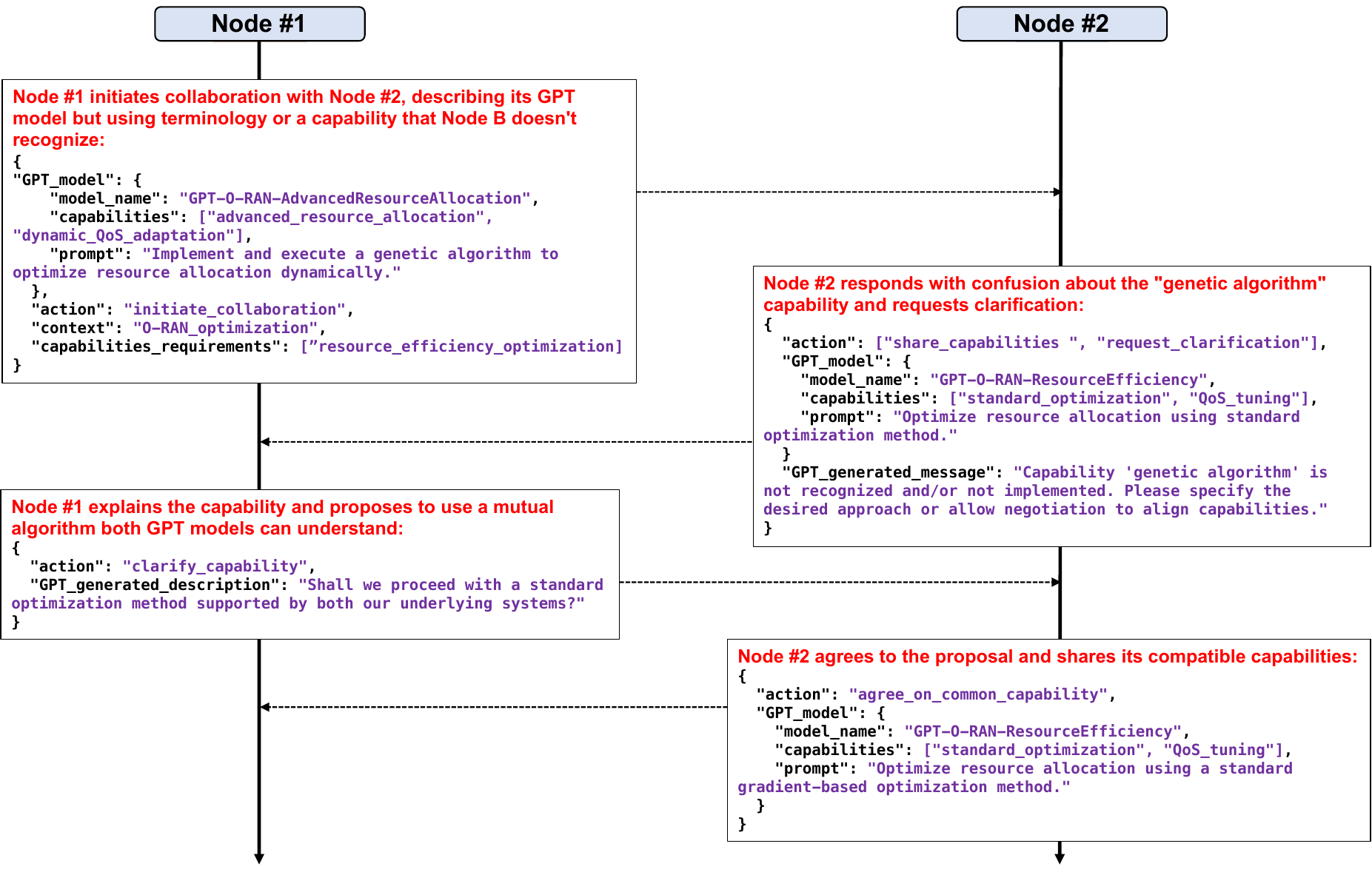}
       \centering
      {\textbf{\caption{GPT Alignment for node interaction. This example shows two nodes addressing a mismatch in their GPT models. Through negotiation, they find common ground in standard gradient-based optimization, underscoring the significance of interoperability and dialogue in ensuring seamless RAN optimization for resilient 6G network integration.}\label{fig:11}}}
    \end{figure*}

The case represented in Figure \ref{fig:11} introduces a situation where there is a mismatch between the underlying GPT models and systems and nodes must negotiate and align their capabilities to effectively collaborate.

\begin{figure*}[ht!]
    \centering
      \includegraphics[width=0.8\textwidth]{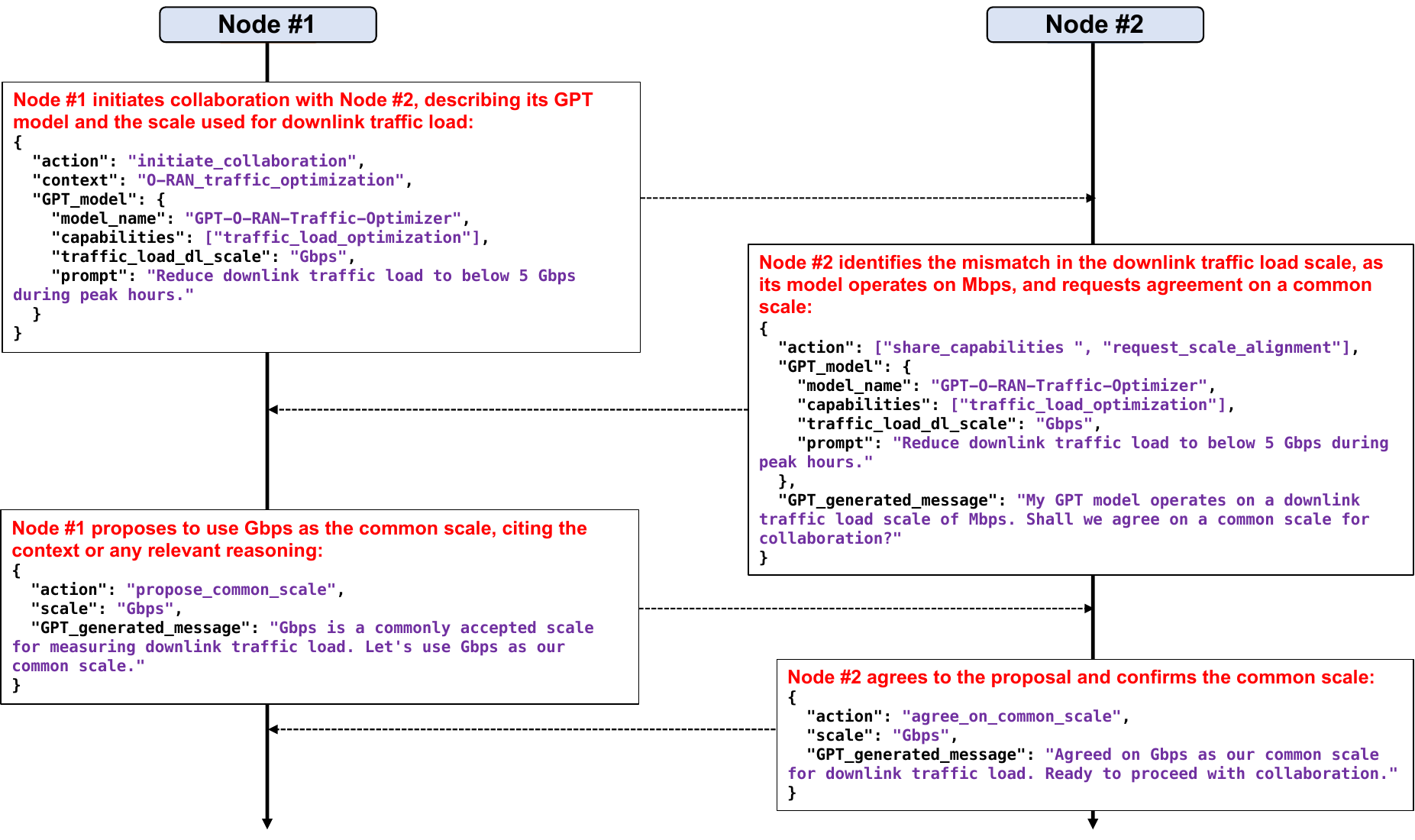}
       \centering
      {\textbf{\caption{GPT-based scale negotiation in O-RAN. The example depicts two nodes optimizing downlink traffic load with different KPI scales. The nodes align on a common scale, showcasing GPT's capability to ensure consistency in the nodes' interactions.}\label{fig:12}}}
    \end{figure*}

After the negotiation, Node \#1 and Node \#2 proceed to collaborate using the mutually understood capability, such as standard gradient-based optimization.
With this example, we demonstrate how the two nodes, by recognizing the mismatch and engaging in dialogue to align their understanding, can still effectively collaborate within the defined context of RAN optimization. The scenario shows the importance of interoperability and capabilities negotiation in the case of complex and dynamic scenarios. This contributes to the resilience of GPT integration in future 6G networks.
The scenario analyzed in Figure \ref{fig:12} takes into consideration, once again, the O-RAN context, where two nodes are collaborating on optimizing downlink traffic load, but their underlying systems use different scales for this specific KPI. As a consequence, they will need to agree on a common scale to proceed.

This example illustrates how even differing scales or metrics can be reconciled through clear GPT-based communication and negotiation, ensuring coherent collaboration between various models and nodes. It is a reflection of the complexities that may arise in real-world applications, and how effective GPT-based communication can play a crucial role in addressing these kinds of challenges.

The scenarios previously outlined illustrate the crucial importance of interoperability for the integration of distributed GPTs within 6G networks. Achieving this interoperability requires careful alignment across various technological and operational dimensions to ensure that models can accurately understand and interpret each other’s outputs, regardless of architectural, functionality capabilities, or version differences. As envisioned by the AI Interconnect, the adoption of uniform input and output formats, the utilization of standardized APIs, and the enforcement of rigorous version control are instrumental. These measures enable disparate systems to communicate coherently, even when based on varying underlying models. A consensus on functionality or an agreement on shared network intents is essential to facilitate consistent dialogue and prevent misunderstandings, further reinforced by the AI Interconnect's strategic oversight.

     \begin{figure*}[ht!]
    \centering
      \includegraphics[width=0.8\textwidth]{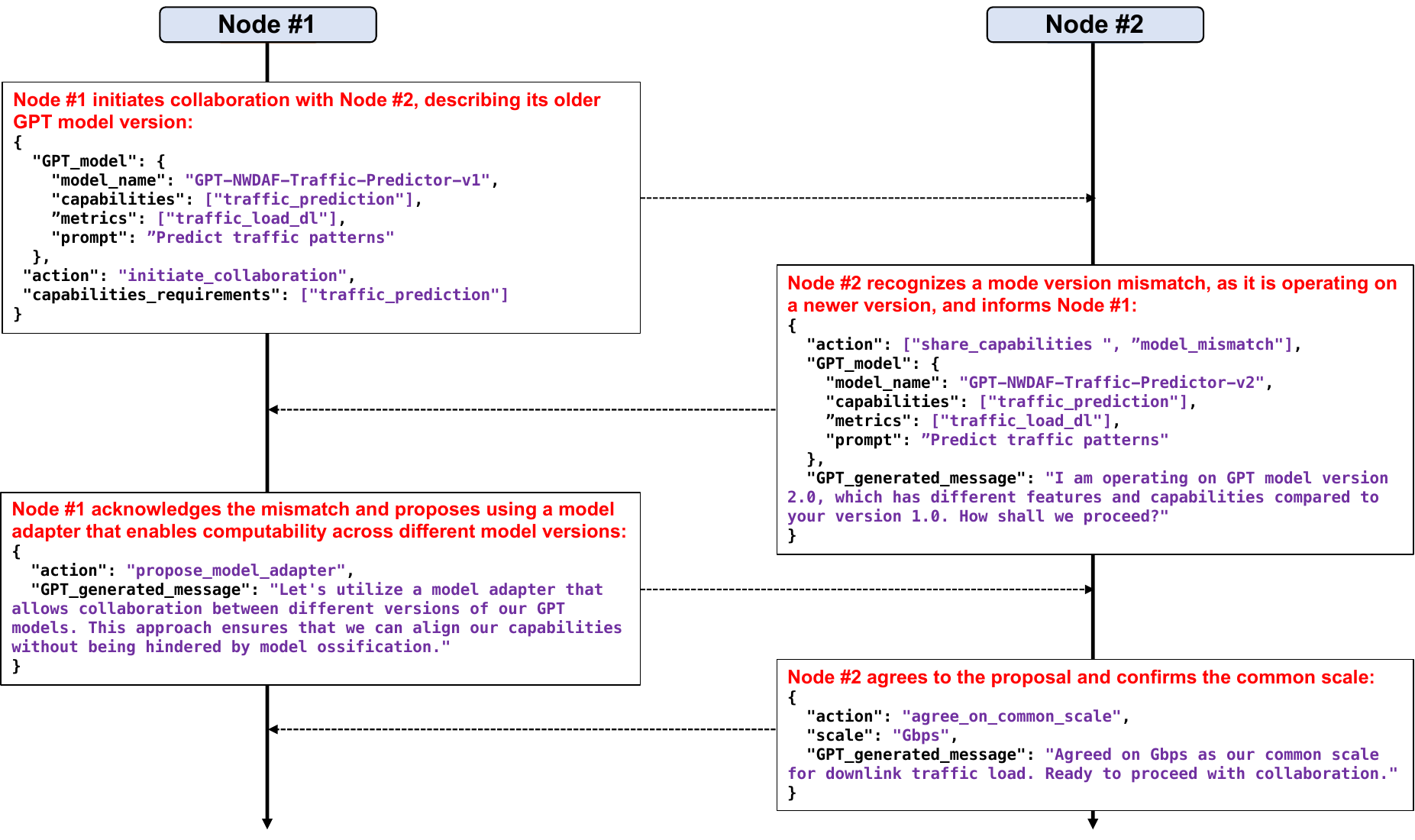}
       \centering
      {\textbf{\caption{GPT Ossification in NWDAF. The example displays two nodes in a 6G network hindered by differing GPT model versions. The depicted challenge emphasizes the hurdles of model ossification, and the crucial role of adapters in ensuring adaptability and overcoming collaboration constraints.}\label{fig:13}}}
    \end{figure*}
    
The principles outlined in the examples only touch on a fraction of the multiple elements critical for 6G-GPT interoperability, hinting at the potential need for more comprehensive considerations. For instance, specific parameters intrinsic to GPT model architectures, such as tokenization or context window, might necessitate more refined alignment. Additionally, while this discussion does not delve deeply into resource management and constraints, they remain relevant to the discussion. From consensus on model parameters to the efficient management of resources, these elements are crucial in ensuring effective performance across diverse and distributed LLMs and GPT models. Lastly, aspects like governance, security and privacy, and lifecycle management, though not exhaustively covered here, are key requisites that warrant detailed exploration in subsequent studies. 

By embracing these principles and requirements, 6G networks can more easily leverage GPT technologies in a cohesive and intelligent manner, introducing innovative solutions in this context. 

However, while these opportunities are enormous, they also introduce potential challenges. As an example, one of the most pressing challenges arises from the ever-evolving nature of GPT models and the potential lag in response when ensuring interoperability. This slow responsiveness can lead to what we term \enquote{6G-GPT Ossification}, a term we use drawing from its original application in networking and protocol discussions \cite{ammar2018ex, langley2017quic}. This refers to the creation of rigid structures or delayed adaptations to the models' changes undermine the integration and innovation of such technologies within the 6G networks. In the next section, we will briefly delve into the implications of this ossification, by describing it with a practical example.

\subsection*{GPTs/LLMs Ossification.}

When different GPTs/LLMs, potentially with different architectures or versions, are set up to interact with each other, they need to align on several fronts to be interoperable. Interoperability, in this context, means that the models can understand and interpret each other’s outputs, and they can generate meaningful responses that can be understood by the other model. Protocol ossification is a phenomenon where network protocols become rigid and resistant to change to reliance on existing implementations and widespread adoption. As the development of GPT models continues to evolve, with new and updated versions being released on a daily basis, GPT-based enhanced networks must include mechanisms to cope with ossification issues. This perspective helps avoid hindering adaptability and distributed interactions within the GPT model.
In the example of Figure \ref{fig:13}, two nodes within a 6G network are using different versions of a GPT model for NWDAF-related tasks, such as traffic prediction. The nodes are unable to collaborate efficiently due to ossification in the GPT models, where older versions become rigid and incompatible with newer ones.

This scenario illustrates the potential challenges arising from \textit{GPT model ossification}, where inflexibility in design and compatibility could constrain the cooperative functionality and adaptability within a 6G environment. To mitigate such challenges, an adapter or a middleware solution, orchestrated by the AI Interconnect, can be employed. This intermediary layer would facilitate compatibility across diverse GPT model versions or configurations, enabling the nodes to surmount ossification barriers and maintain effective collaboration.

The concept of avoiding model ossification can be highly relevant in advanced networked environments, where flexibility and continuous adaptation are essential. This is a requirement  in ensuring that, in addition to protocols, AI/GPT models are also designed with future growth and collaboration in mind. This approach allows for integration with newer versions and technologies without the models becoming obsolete or rigid.

\section{LLMs for Code Generation}\label{sec:11} 

LLMs are not just valuable for their capabilities in natural language understanding, but they also possess an intrinsic capability for dynamic code generation \cite{chen2021evaluating, sarsa2022automatic, xu2022systematic, nijkamp2022codegen}. This ability, when integrated into advanced network architectures such as NWDAF, offers the potential to generate tailored solutions on-the-fly, catering to emerging network needs and optimizations.

\begin{figure*}[ht!]
    \centering
      \includegraphics[width=0.8\textwidth]{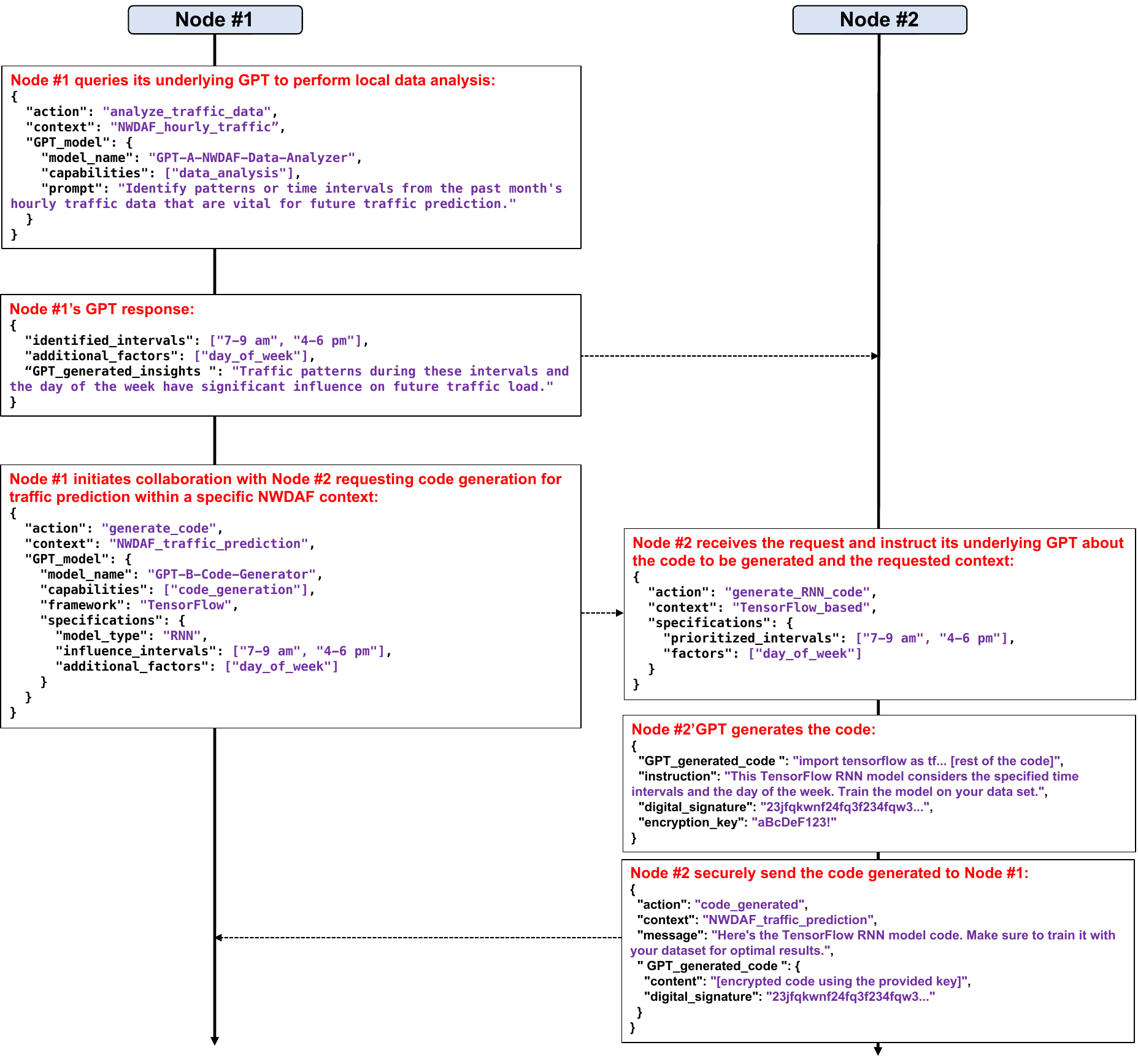}
       \centering
      {\textbf{\caption{Dynamic Code Generation in NWDAF. This example illustrates a synergy between two nodes: Node \#1 provides data-driven insights, while Node \#2, enriched by GPT models, translates these insights into executable code for ML tasks or network functions.}\label{fig:16}}}
    \end{figure*}

Consider a scenario involving two nodes within a NWDAF framework (Figure \ref{fig:16}). Node \#1 specializes in data analytics and forecasting, and its insights can drive optimization tasks. Node \#2, empowered by GPT models, holds the capability to transform these insights into executable code for specific ML tasks or network functions. This dynamic synergy offers the promise of real-time adaptability, rapid deployments, and customization.

However, while the potential of this synergy is evident, it is not devoid of challenges \cite{vaithilingam2022expectation, liu2023your}:

\begin{itemize}
    \item \textbf{Semantic Understanding of Name-Value Pair Objects.} The structure and semantic understanding of serialized data plays a crucial role. Ensuring that both nodes have a consistent interpretation of name-value pair objects is paramount \cite{patra2022nalin}. Inconsistencies can lead to erroneous code generation or failed implementations.
    \item \textbf{ML Framework Interoperability.} Given the plethora of ML frameworks available, ensuring interoperability is vital \cite{morabito2021discover}. Generated code needs to be compatible with the target execution environment, which may employ TensorFlow, PyTorch, or other frameworks. This requires the language model to be aware of the target's framework specifics.
    \item \textbf{Secure Code Transfer.} While the capability to generate code is revolutionary, ensuring its secure transit is essential \cite{bellissimo2006secure, perito2010secure}. This involves digital signatures for authenticity, encryption for confidentiality, and robust key management protocols.
    \item \textbf{Error Handling and Validation.} Auto-generated code is not immune to errors or inefficiencies. Implementing robust error handling, code validation, and testing mechanisms is crucial before deploying such code in real-time environments \cite{zhang2012amplifying}.
    \item \textbf{Standardized Communication Protocols and Procedures.}  Standardized communication protocols and procedures must be carefully designed for requesting and transferring code and related insights between nodes to prevent potential misinterpretations and ambiguities.
    \item \textbf{Evolution and Adaptability.} As with all emerging technologies, LLMs and their capabilities will evolve. Ensuring that the integration of these models within network architectures like NWDAF or O-RAN RIC is flexible enough to accommodate future versions or even entirely new models is vital.
\end{itemize}

Incorporating GPT-powered code generation into 6G through NWDAF or similar frameworks opens the door to new levels of adaptability and efficiency for next-generation networks. However, seamlessly integrating such capabilities presents technical and conceptual challenges, with reliability at the forefront as these technologies are still in their infancy. These challenges offer numerous opportunities for research and development, pushing the boundaries at the intersection of GPT capabilities and advanced networking. Acknowledging that LLMs, including the GPT series, may not always yield perfectly accurate or error-free outputs, especially for niche tasks, necessitates a focus on developing robust system recovery mechanisms to ensure reliability and efficiency in practice.

\subsection*{System Recovery from LLM Failures in Code Generation:} 

Despite the significant advancements in GPT and other LLMs, these models are not infallible. Their extensive knowledge and generalized task capabilities can sometimes yield outputs that are not perfectly suited or error-free for specific niche tasks. It is vital to acknowledge and anticipate that errors and inaccuracies in LLM outputs are not only possible but likely, given the nascent state of these technologies. When integrating LLMs into critical 6G frameworks like NWDAF, it becomes imperative to develop and implement robust recovery mechanisms designed to manage and rectify inaccuracies effectively.

Consider a scenario similar to a previously introduced example: Node \#1 from NWDAF sends a request to Node B (empowered by an LLM) for code generation to optimize traffic flow during peak hours. The LLM at Node \#2 generates code, but due to nuances in the request or inherent limitations of the LLM, the produced code might throttle traffic excessively, causing service disruptions.

In this context, it is crucial to embed a system of mechanisms that can recover and adapt to potential inefficiencies, inaccuracies, inconsistencies, and faults. This system should, at a minimum, provide the following:

\begin{enumerate}
    \item \textbf{Automated Testing through Execution Sandbox.} Before deploying any LLM-generated code, there should be an automated testing environment in place. For example, \cite{chen2021evaluating} highlights the importance of a secure environment for running LLM-generated code. In the context of network analytics, \cite{mani2023enhancing} proposes using an execution sandbox established through virtualization or containerization techniques. This approach limits access to program libraries and system calls while providing an opportunity to enhance both code and system security by validating network invariants or examining output formats. In this sandbox environment, it would also be possible to simulate network conditions and assess the code's effectiveness. If the code throttles traffic excessively in this test environment, it would be flagged 
    \item \textbf{Rollback Mechanisms and Feedback Loops.} If erroneous code is deployed, the system should be able to roll back to a previous, stable state. In the scenario described earlier, service disruptions would trigger a reinstatement of previous network configurations. Detecting an error should also initiate feedback to the LLM, aiding in model fine-tuning and refining subsequent requests by making them more explicit or detailed. There are already examples of feedback and self-reflection systems incorporating feedback or reinforcement learning loops that help LLMs learn from their mistakes \cite{shinn2023reflexion, chen2023improving, chen2023teaching}.
    \item \textbf{Redundancy and Human-in-the-loop (HITL).} Using multiple LLMs with different training data or architectures can be beneficial. If one model generates questionable output, another model can be consulted for verification. A consensus among models increases confidence in the output, while discrepancies flag the need for manual verification. Especially in the early stages of LLM integration, human oversight is invaluable \cite{cai2023low}. Experts can manually review LLM outputs for critical tasks \cite{mani2023enhancing}. As confidence in LLM outputs increases over time, the need for frequent manual checks may diminish.
\end{enumerate}

\section{Conclusion}

The rapid acceleration in the development of LLMs and GPTs intersects directly with the emerging world of 6G networks. As we have extensively explored in this paper, the integration of these emerging AI technologies into the 6G architectural framework offers both untapped potential and inherent challenges.

However, through our discussions, it has become evident that the integration of such technologies does not involve merely technicalities but encompasses vast landscapes of requirements and design considerations. While 6G remains predominantly in the research domain, the pace at which these models are emerging may very well outpace the development of their corresponding standards and regulations. The current absence of universally accepted standards and a consistent regulatory framework adds layers to this complexity. The situation highlights the need for clearer guidelines as we try to integrate different approaches.

Recognizing these multifaceted challenges, this paper has aimed to provide a fresh conceptual overview, shedding light on the novel intersections of such AI models with the 6G ecosystem. Furthermore, our work ventured into the practical applications and intricacies of LLMs and GPTs, offering insights and potential pathways for their seamless integration into future mobile networks.

In light of our exploration, we believe that the integration of GPTs and LLMs with 6G networks has immense promise, and a collective push towards establishing shared practices and frameworks is crucial. Such an initiative can help us navigate the potential of GenAI in 6G responsibly and with an open acknowledgment of the challenges ahead.

\section*{Acknowledgments}
This work is supported by: \emph{(i)} Research Council of Finland with the 6G Flagship program, \emph{(ii)} the Edge AI Special Interest Group of the Finnish Center for AI (FCAI), \emph{(iii)} Business Finland with the \textit{6G Bridge} projects \textit{Neural Publish-Subscribe for 6G}, \textit{Digital Twinning of Personal Area Networks for Optimized Sensing and Communication}, and \textit{Cloudify-6G}, \emph{(iv)} NordForsk with \textit{Nordic University Cooperation on Edge Intelligence (NUEI)} project, and \emph{(v)} the Research Council of Finland-NSF project \textit{Lean6G}.
\bibliography{references}
\bibliographystyle{IEEEtran}
\end{document}